\documentclass[useAMS,usenatbib]{mn2e}

\usepackage{graphicx}
\usepackage{hyperref}
\usepackage{multirow}

\def\gse{\mathrel{\mathchoice {\vcenter{\offinterlineskip\halign{\hfil
$\displaystyle##$\hfil\cr>\cr\simeq\cr}}}
{\vcenter{\offinterlineskip\halign{\hfil$\textstyle##$\hfil\cr
>\cr\sim\cr}}}
{\vcenter{\offinterlineskip\halign{\hfil$\scriptstyle##$\hfil\cr
>\cr\sim\cr}}}   
{\vcenter{\offinterlineskip\halign{\hfil$\scriptscriptstyle##$\hfil\cr
>\cr\sim\cr}}}}}

\newcommand{\arxiv}[1]{}

\newcommand{\arxivpreprint}[1]{(\href{http://arxiv.org/abs/#1}{arXiv:#1})}

\title{30~GHz observations of sources in the VSA fields}
\author[M.\,P. Gawro\'nski et al.]
 {M.\,P.~Gawro\'nski$^1$, M.\,W.~Peel$^2$, K.~Lancaster$^3$, R.\,A.~Battye$^2$, M.~Birkinshaw$^3$, 
 \newauthor I.\,W.\,A.~Browne$^2$, M.L.~Davies$^4$, R.\,J.~Davis$^2$, R.~Feiler$^1$, T.\,M.\,O.~Franzen$^4$,
 \newauthor R.~G\'enova-Santos$^4$, A.\,J.~Kus$^1$, S.\,R.~Lowe$^2$, B.\,M.~Pazderska$^1$, E.~Pazderski$^1$,
 \newauthor G.\,G.~Pooley$^4$, B.\,F.~Roukema$^1$, E.\,M.~Waldram$^4$ and P.\,N.~Wilkinson$^2$\\
 $^1$ Toru\'n Centre for Astronomy, Nicolaus Copernicus University, 87-100 Toru\'n/Piwnice, Poland \\
 $^2$ Jodrell Bank Centre for Astrophysics, The University of Manchester, Manchester, M13 9PL\\
 $^3$ University of Bristol, Tyndall Avenue, Bristol, BS8 ITL\\
 $^4$ Astrophysics Group, Cavendish Laboratory, JJ Thomson Avenue, Cambridge, CB3 0HE}
 
\begin{document}

\date{Accepted \bf [2009 Month DD]\rm. Received \bf [2009 Month DD]\rm; in original form \bf [2009 Month DD] \rm}

\pagerange{\pageref{firstpage}--\pageref{lastpage}} \pubyear{2009}

\maketitle

\label{firstpage}

\begin{abstract}
Small angular scale (high $\ell$) studies of cosmic microwave background anisotropies require accurate knowledge of the statistical properties of extragalactic sources at cm-mm wavelengths. We have used a 30~GHz dual-beam receiver (OCRA-p) on the Toru\'n 32-m telescope to measure the flux densities of 121 sources in VSA fields selected at 15~GHz with the Ryle Telescope. We have detected 57 sources above a limiting flux density of 5~mJy, of which 31 sources have a flux density greater than 10~mJy, which is our effective completeness limit. From these measurements we derive a surface density of sources above 10~mJy at 30~GHz of $2.0~\pm~0.4$ per square degree. This is consistent with the surface density obtained by \citet{Mason2009} who observed a large sample of sources selected at a much lower frequency (1.4~GHz).  We have also investigated the dependence of the spectral index distribution on flux density by comparing our results with those for sources above 1~Jy selected from the WMAP 22~GHz catalogue. We conclude that the proportion of steep spectrum sources increases with decreasing flux density, qualitatively consistent with the predictions of \citet{deZotti2005}. We find no evidence for an unexpected population of sources whose spectra rise towards high frequencies, which would affect our ability to interpret current high resolution CMB observations at 30~GHz and above.

\end{abstract}


\section{Introduction} \label{sec:introduction}

Extragalactic radio sources are a serious contaminant in observations of primordial Cosmic Microwave Background (CMB) anisotropies on small angular scales (high $l$; e.g. \citealp{Toffolatti2005}).  In order to estimate the CMB temperature power spectrum with high accuracy, detailed information about the foreground population is required.  CMB experiments can be designed to work at any of a range of frequencies from $\sim$20 to $\sim$300~GHz, since the spectrum peaks at $\sim160$~GHz. Extragalactic radio sources are problematic over most of this range, with different source populations, from active galaxies to star-forming galaxies, dominating the contamination in different wavebands. However, information on these contaminating populations is limited, since large-scale high-resolution surveys of the radio sky are restricted to lower frequencies, as for example in the NRAO VLA Sky Survey \citep[NVSS;][]{Condon1998} at 1.4~GHz and the Green Bank 4.85~GHz survey \citep[GB6;][]{Gregory1996}.  Model-dependent predictions of the contaminating sources at the frequencies of CMB experiments may be made \citep[e.g.][]{Toffolatti1998,deZotti2005}, but these can be unreliable. Source populations can exhibit a wide range of spectral indices so that severe selection effects may arise when extrapolating to high frequencies from the limited set of surveys on which these predictions are based.  Thus, for maximum reliability in their cosmological interpretation, it is essential that CMB observations are complemented by deep and carefully-selected high-frequency radio source surveys.

The CMB power spectrum has been measured over a wide range of angular scales, from the early low-resolution (7$^\circ$ per beam) work of the {\it Cosmic Background Explorer} \citep[COBE,][]{Smoot1992}, through the intermediate scales probed by e.g. the Very Small Array (VSA, see e.g. Dickinson et al. 2004), {\sc Boomerang} \citep[see e.g.][]{Jones2006} and the {\it Wilkinson Microwave Anisotropy Probe} \citep[WMAP, see e.g.][]{Hinshaw2003,Hinshaw2009}, up to the high resolution measurements of e.g. the Cosmic Background Imager \citep[CBI, see e.g.][]{Sievers2009} and the Arcminute Cosmology Bolometer Receiver Array \citep[ACBAR, see e.g.][]{Reichardt2009}. The problem of contamination by radio sources increases at high multipole order, $\ell$, since the power spectrum of point sources increases as $\ell^2$.  Source subtraction strategies are therefore becoming ever more important. This is particularly so because of the results  obtained with the CBI at 30~GHz, which indicate an excess of power at high $\ell$ compared to best fit cosmological models, the so-called `CBI excess' \citep{Mason2003,Readhead2004,Sievers2009}. There has been much discussion of this result with some new observations not detecting a significant excess of power at  $\ell >2000$ \citep{Sharp2009,Friedman2009,Reichardt2009}, though most either do not cover the same range of multipoles or are at different observing frequencies.

One possibility is that the excess arises from a population of point sources, another is that that the power comes from the Sunyaev-Zel'dovich (SZ) effect associated with clusters of galaxies. \citet{Sievers2009} interpret the CBI excess as due to the SZ effect, provided that the value of $\sigma_8$, the normalization of the power spectrum of the density fluctuations in the Universe on scales of $8 h^{-1}$~Mpc, is significantly higher than that calculated from WMAP data \citep{Komatsu2009}. However, a value of $\sigma_8$ consistent with the WMAP results may still be allowed when the increased variance and skewness of the SZ effect statistics within small areas of the sky are taken into account \citep{Peel2009,Shaw2009}. The issue of point source contamination has been addressed by \citet{Sievers2009} using 30~GHz measurements of NVSS sources in the CBI fields made with the GBT by \citet{Mason2009}. \citet{Sievers2009} state that source contamination cannot account for the excess. The GBT 30~GHz sample was, however, selected at 1.4~GHz from the NVSS survey, so it is possible that a population of sources with spectra rising towards higher frequencies could have been missed. 

In order to look for the CBI excess, the VSA was upgraded and reconfigured in 2005 to provide both higher sensitivity and higher resolution, becoming the VSA `Super-Extended Array'. With the issue of point source contamination in mind from the outset, the VSA system includes a separate pair of telescopes operating as an interferometer for source subtraction. As the Super-Extended fields lie
within those observed during the project's earlier phases (see \citealp{Dickinson2004} and \citealp{Grainge2003}), information is therefore available on the stronger ($\sim 20$ mJy) sources in the VSA target fields from the 33~GHz flux density measurements made using these telescopes \citep{Cleary2005}. However, as discussed by \citet{Sievers2009}, sources at the mJy level could combine to produce an excess of power at high multipoles. For this reason it is important to extend the work of \citet{Cleary2005} to lower flux densities, however this could not easily be achieved using the VSA source subtractor. Ideally one would like to carry out a 30~GHz blind survey of sources down to flux densities of a few mJy but this is technically very challenging so an alternative approach is required.

In this paper, we present the results from a collaboration between the VSA and the One Centimetre Receiver Array (OCRA, see e.g. \citealp{Browne2000}; \citealp{Lowe2007}; \citealp{Lancaster2007}; \citealp{Pazderska2009}) teams to study the population of radio sources that affect deep VSA measurements. The VSA group have surveyed the CMB target fields at 15~GHz using the Ryle Telescope \citep[RT, see][]{Waldram2003}. We have then used the OCRA-p receiver, a 30~GHz dual-beam system, sensitive to a single linear polarization, mounted on the 32-m Toru\'n telescope to measure the flux densities of the RT sources at 30~GHz, which is similar to the VSA's operating frequency of 33~GHz.

Studies at 30\,GHz of sources selected at 15\,GHz also provide new information about the poorly-studied properties of radio source populations at high radio frequencies. Understanding these populations at higher frequencies is vital for upcoming work, not only for studies of CMB primordial anisotropies, but also planned cluster surveys utilising the Sunyaev-Zel'dovich effect.

\section{The 15~GHz sample from the RT survey} \label{sec:ryle}

\subsection{The observations}

The five fields which have been observed by the VSA Super-extended Array, designated VSA 1L, 2G, 3L, 5E, 7E, are centred at RA($^{\rm h}\ ^{\rm m}\ ^{\rm s}$), Dec($^\circ\ ^\prime\ ^{\prime\prime}$) J2000: 00 16 09.1, +28 16 40 (1L); 09 40 52.7, +31 46 21 (2G); 15 38 35.3, +41 40 17 (3L); 03 02 57.0, +26 11 44 (5E); 12 32 21.8, +52 43 27 (7E). These areas lie within those surveyed earlier by the RT for the work of \citet{Cleary2005}, but the central regions, out to a radius of 1$^\circ$, have now been scanned more deeply. This radius corresponds approximately to the first null in the relevant VSA primary beam.

Since the FWHM of the primary beam of the RT is very small compared with that of the VSA (6~arcmin for the RT and 72~arcmin for the VSA), we have again used a rastering technique similar to that explained in \citet{Waldram2003}.  The raster observations are combined to form so-called raster maps which are used to identify peaks as possible sources. Each candidate is followed up with a short pointed observation, either to establish a reliable flux density or to eliminate it as a false detection. More details of how we have developed our method to reach greater sensitivity appear in \citet{Waldram2009}, which describes the deeper 9C 15~GHz survey.

Each field has a sky area of 3.14~deg$^{2}$, apart from 3L in which a very small fraction of the field ($<1$ per cent), close to its low declination limit, was not covered by the survey. The total area for all five fields amounts to 15.69~deg$^{2}$ or 0.004778~sr. We have not assigned individual error estimates to the source postions and flux densities. The positions are derived from the raster maps and, by examining a sample of sources with independent measurements on overlapping maps, we conclude that the positional accuracy is better than 10~arcsec. The flux densities are from the pointed observations and the uncertainty in these is dominated by the random uncertainty in the flux calibration, which we estimate to be $\sim5$~per cent. 

\subsection{Completeness} \label{sec:completeness}

\begin{figure}
\centering
\includegraphics[width=6.5cm]{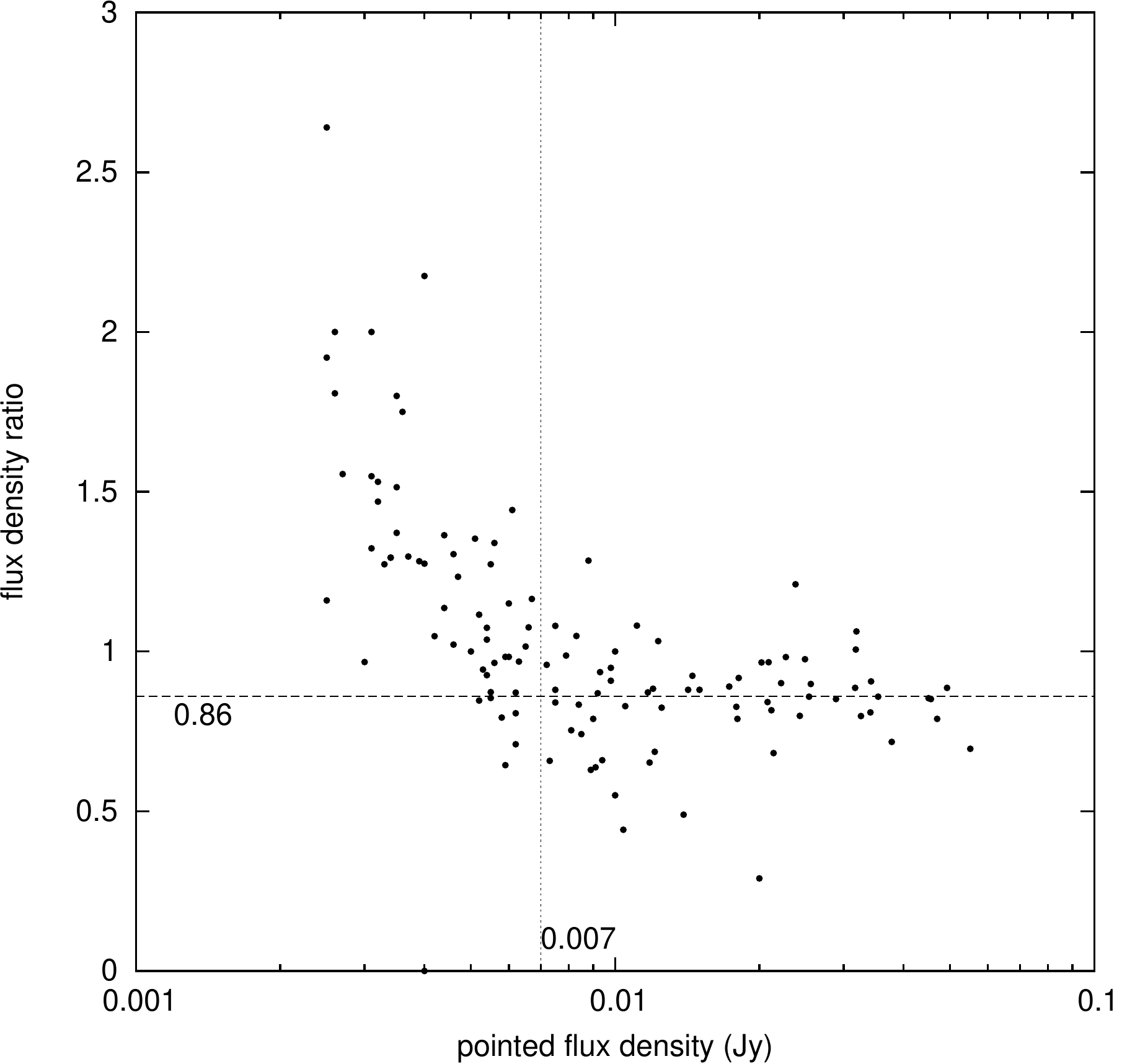}
\caption[]{Plot of the ratio $(S_{\mathrm{r}}/S_{\mathrm{p}})$ of raster flux density to pointed flux density versus $S_{\mathrm{p}}$. The estimated completeness limit is $\approx7$mJy.}
\label{fig:fratio}
\end{figure}

\begin{figure}
\centering
\includegraphics[width=9cm]{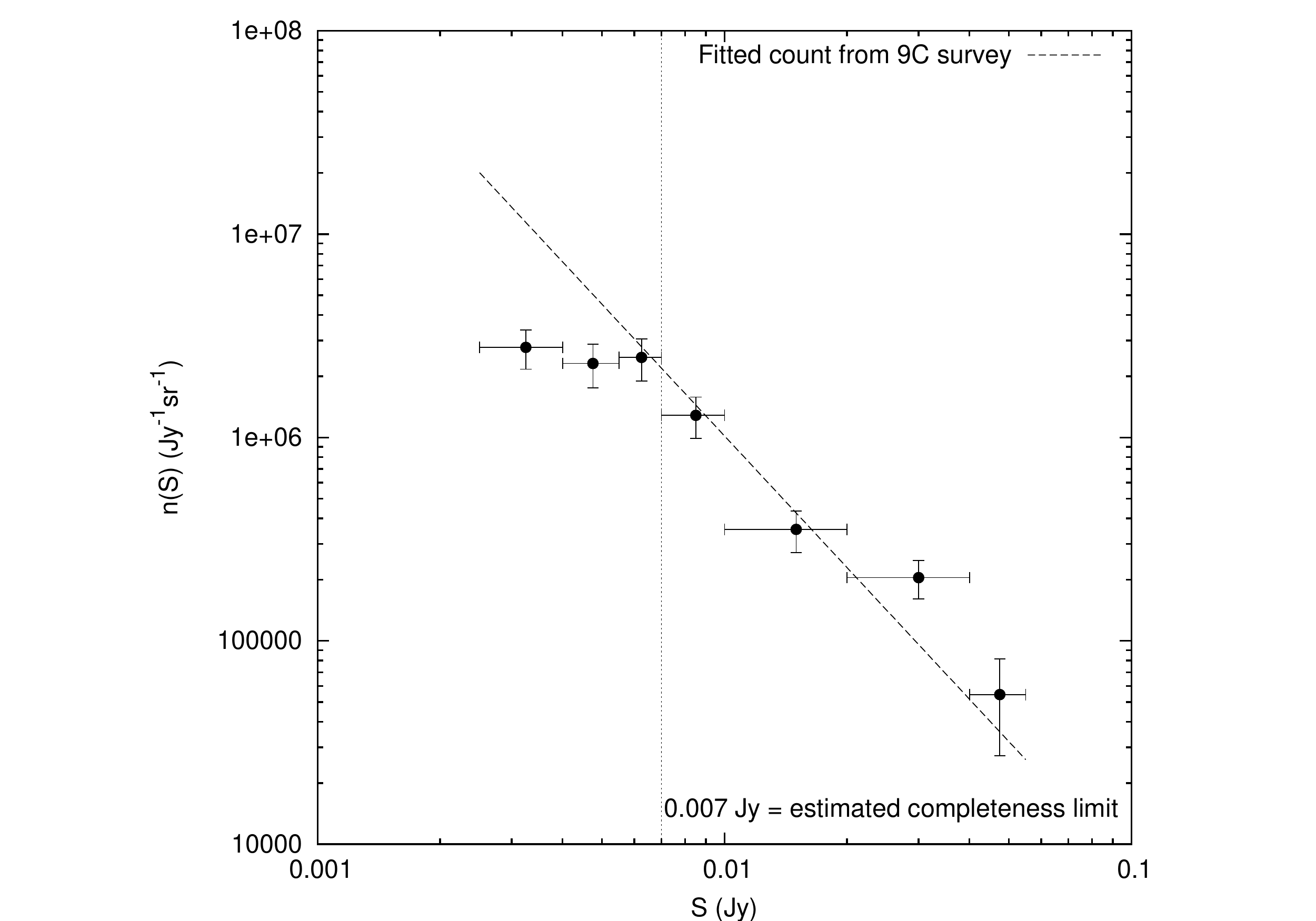}
\caption[]{The differential source count for the 15~GHz sample, showing the estimated completeness limit and the fitted 15~GHz count from the 9C survey.}
\label{fig:count}
\end{figure}

For the purpose of this paper an important question to consider is the completeness of the 15~GHz sample. The raster maps used to identify the candidate sources have a range of noise levels, from as low as $<1$~mJy to as high as 1.3 or 1.4~mJy at the edges of some of the fields. This means that, although we can be sure  that every source in the catalogue is genuine, since each one has been followed up with a pointed observation, we are less sure of the number of undetected sources at any flux density level. We have approached this problem by a method described in detail in the deeper 9C paper \citep{Waldram2009}. Basically, we plot the ratio of the raster to the pointed flux density, $(S_{\mathrm{r}}/S_{\mathrm{p}})$, versus $S_{\mathrm{p}}$ (see Figure \ref{fig:fratio}). At the higher flux densities, where we can assume the sample to be complete, this ratio has a median value of 0.86 (see the above paper for an explanation of why it is less than unity) but for the weaker sources the ratio rises systematically with decreasing flux density. This is because at these levels the peaks detected on the raster maps are preferentially those boosted by a positive local noise contribution. By investigating the progressive increase in the ratio with decreasing pointed flux density, we conclude that our completeness limit is approximately 7~mJy. This is consistent with the differential source count for the sample (see Figure \ref{fig:count}) which shows a significant fall-off compared to the 9C survey single power-law fit at the lower flux densities. There are 65 sources above the completeness limit and 56 below.

The completeness limit of the 15~GHz sample has implications for the 30~GHz list. On the basis that very few sources have a rising spectral index $\alpha_{15}^{30} \geq 0.5$ where $S_\nu \propto \nu^{\alpha}$, we conclude that we should have detected all sources within the VSA fields stronger than 10~mJy at 30~GHz. Hence our list of sources stronger than 10~mJy forms a complete statistical sample at 30~GHz.

\section{OCRA observations and calibration methodology} \label{sec:observations}
\begin{figure}
\centering
\includegraphics[scale=0.34]{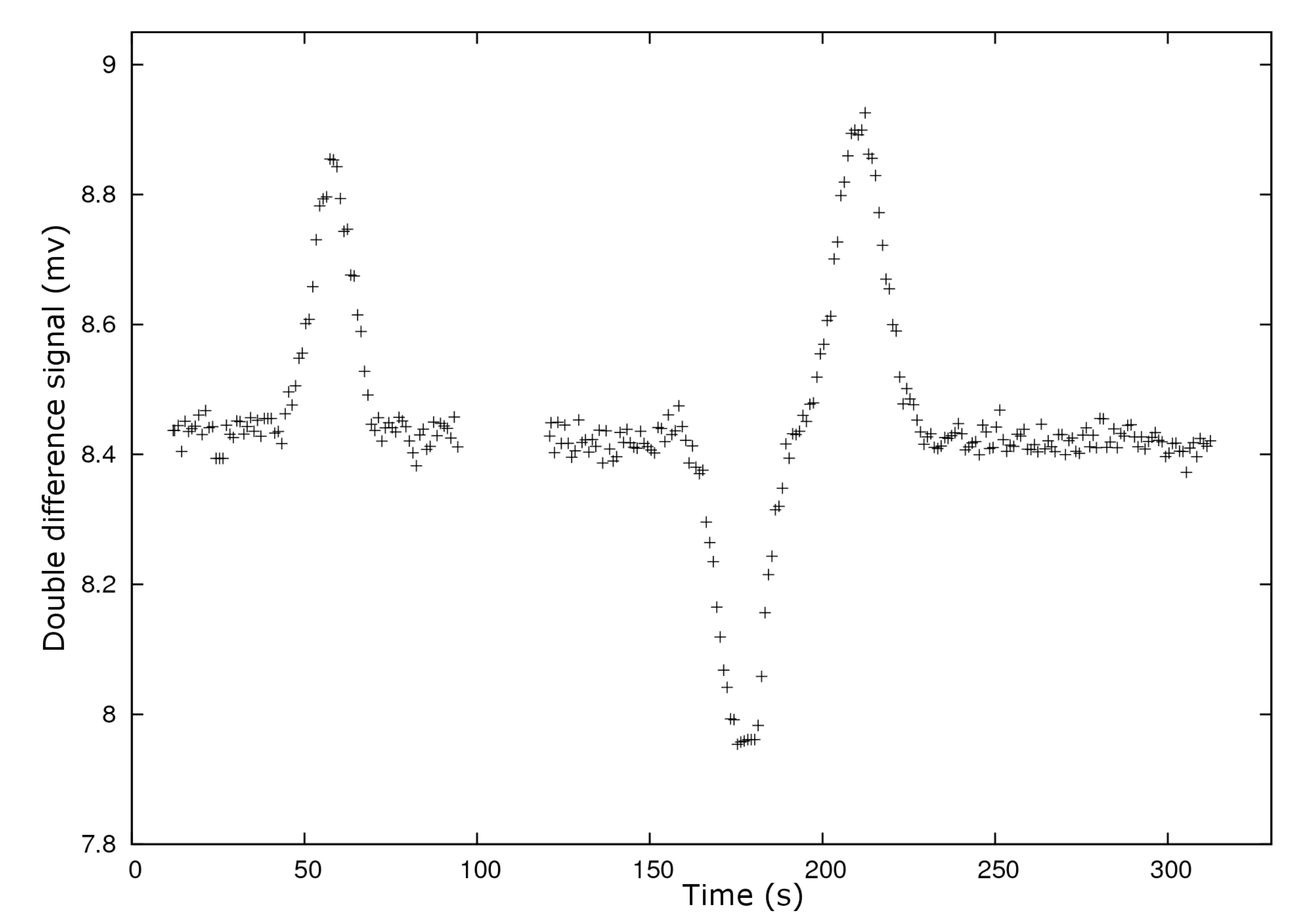}
\caption[]{A cross-scan measurement of J0958+3224. First, a scan of the source is made in elevation, during which one beam crosses over the source. A Gaussian function is immediately fitted to this scan and an initial pointing correction is calculated and applied. A scan in azimuth is then performed, where both beams cross over the source. This is fitted using a double Gaussian function, and the final pointing correction is calculated. The initial pointing error is easily noticeable as there is a difference in heights of the beams in the two scans. The one second samples from these scans are shown.}
\label{fig:qscan}
\end{figure} 
\begin{figure}
\centering
\includegraphics[scale=0.34]{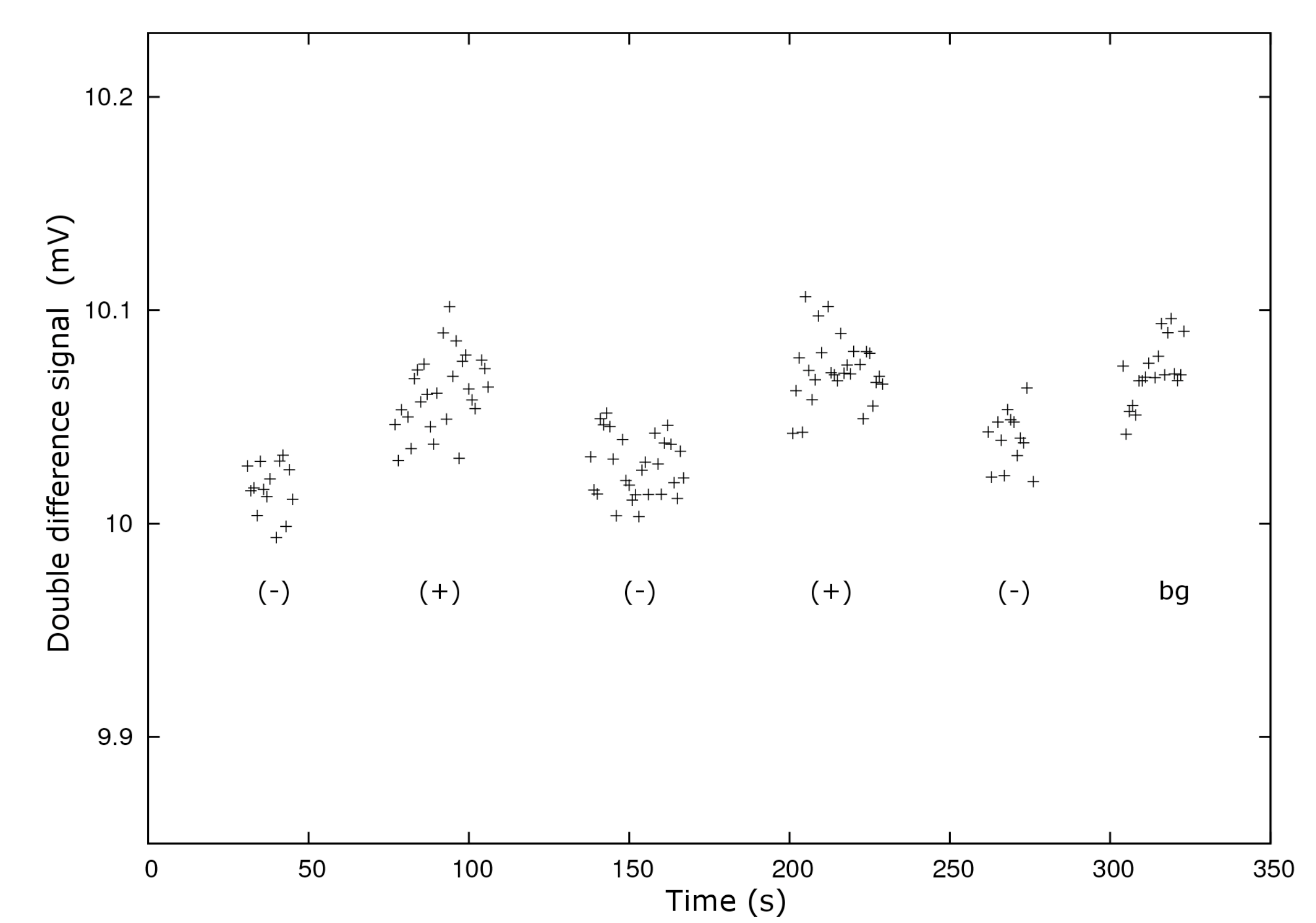}
\caption[]{An example of an on-off observation on J1538+4105 (27.6 $\pm$ 1.2 mJy). The first integration is taken when the source is tracked by the secondary beam (marked as (-) on the plot) for 15 seconds (half of the standard integration time of 30~s). Next, three integrations are made, alternating between the primary and secondary beam, followed by a final integration with the source in the secondary beam of the same duration as the first. The last step is an integration at a background position 0.5 deg away in azimuth (marked as `bg').}
\label{fig:onoff}
\end{figure}

OCRA-p was used to observe the RT sample between February and October 2007, with additional measurements made during 2008 and early 2009 to complete the sample and check for variability. The pointing positions RT, accurate to better than 10~arcsec, are more than adequate for the 1.2~arcmin OCRA-p beam. Observations of discrete sources are made using two methods. Bright sources (S$_{30\,\mathrm{GHz}} \gse 100$mJy) can be measured using cross-scan observations, where a scan across the source is made first in elevation and then in azimuth (see Figure \ref{fig:qscan}). This method is used both to measure the flux density of target sources \citep[see e.g.][]{Lowe2007} as well as to determine the required pointing corrections for the telescope (discussed further below). As the sources in the VSA fields are generally too weak for the cross-scan technique, a second method was used whereby the respective two beams, the primary `on' beam and the secondary `off' beam, are pointed at the source alternately with a switching cycle $\sim$50 sec for a period $\sim$5 min, thus measuring the source flux density relative to the background on each side of the source. This  method is also used for observing galaxy clusters with OCRA-p (see \citealp{Lancaster2007}). The switching pattern for this on-off method has evolved during the observations and the final version is shown in Figure \ref{fig:onoff}.

The planetary nebula NGC\,7027 was used as our primary flux density calibrator, and was observed at least once per day during the observations of the VSA sources. NGC~7027 is the brightest planetary nebula in the radio sky. \citet{Hafez2008} report a flux density of $5.39 \pm 0.04$~Jy at an epoch of 2003.0 at 33~GHz with a secular decrease of $-0.17 \pm 0.03$ per cent per year. The flux density at the epoch of 2008.0 is therefore $5.34 \pm 0.04$~Jy. We calculate a flux density of $5.47 \pm 0.04$~Jy at 30~GHz, using a spectral index of -0.119 \citep{Hafez2008}, and use this to calibrate all of the 30~GHz flux measurements presented in this paper.  \citet{Zijlstra2008} have also obtained accurate data for the flux density of NGC~7027 from which we obtain a value at 30~GHz of $5.37 \pm 0.28$~Jy at the epoch of 2008.0, which is consistent with the \citet{Hafez2008} result.

Pointing calibrations are carried out every 30-45 minutes (or every 5-6 on-off measurements). The calibration sources used were J0015+3216 (for field 1L), J0958+3224 (2G), J1521+4336 (3L), J0245+2405 (5E) and J1219+4829 (7E).

\section{Data reduction} \label{sec:datareduction}
As the observations are carried out in two different ways -- cross-scans across the calibrators and on-offs for the sources -- the measured flux densities of the sources are found from two different fitting routines. For the azimuthal cross-scans of the calibration sources, two Gaussian peaks and a quadratic background were fitted, and the average of the two peak heights is used.

The on-off measurements of the VSA sources are fitted by taking a symmetric double difference of the data. In this, the average of an `on' measurement $S_{\mathrm{on}}$ has the average of the second half of the preceding `off' measurement $S_{\mathrm{off,prev}}$ and the first half of the following `off' measurement $S_{\mathrm{off,next}}$ subtracted from it, i.e.
\begin{equation}
S_\mathrm{dd} = \frac{1}{N} \sum 0.5 \left(S_{\mathrm{on}} - 0.5 (S_{\mathrm{off,prev}} + S_{\mathrm{off,next}}) \right),
\end{equation}
where the sum is over all $N$ independent off-on-off sets in the data. As the difference between the measured fluxes is twice the source flux, an extra factor of 0.5 is introduced. The internal error on the measurement is calculated by
\begin{equation} \label{eq:error}
\sigma_\mathrm{dd} = \frac{1}{\sqrt{N}} \sum \sqrt{ \left( \frac{\sigma_{S_{\mathrm{on}}}}{2} \right)^2 + \left (\frac{\sigma_{S_{\mathrm{off,prev}}}}{4} \right)^2 + \left( \frac{\sigma_{S_{\mathrm{off,next}}}}{4} \right)^2}.
\end{equation}
Prior to this, any 1 second data points greater than 3 sigma away from the average -- representing spikes in the data -- are removed.

A noise diode is used for secondary flux density calibration. The ratio of outputs obtained with cross-scans of NGC~7027 and the noise diode are averaged for a 24 hour period; this diode calibration is then applied to the individual measurements.

The change of the telescope gain with elevation was corrected for using a fit to observations of NGC~7027 made over a wide range of elevations. As described by \citet{Lowe2007}, atmospheric absorption is corrected for using measurements of the system temperature at the zenith and 60 degrees from zenith made at regular intervals during the observations. A simple `flat earth' model in which the atmosphere is assumed to be a continuous slab with constant thickness and optical depth was used.

A significant number of measurements had to be discarded due to adverse effects of poor weather. Each source was observed multiple times, with individual measurements with large fit errors removed, giving an average of 8.3 observations per source. The final flux density for the source was calculated using the $1/\sigma_{\mathrm{dd}}^2$ weighted mean and corresponding standard error on the mean. To ensure the maximum reliability of the results, the observations were reduced separately at the three institutions involved in the OCRA collaboration, with the resulting values compared and discrepancies identified and resolved. The flux densities given in the next section are from one of these reductions; the other reductions give essentially the same values.

The combined uncertainty from these factors is $\sim 8$ per cent. The systematic error due to the error in the flux calibrator, NGC~7027, is negligible at 1 per cent. As such, the final errors on the 30~GHz flux densities are calculated by $\sigma = \sqrt{\sigma_\mathrm{dd}^2 + (0.08 S)^2}$.

\section{Results} \label{sec:results}
\begin{table*}
\caption{Sources in the VSA fields. The positions are from the RT 15~GHz observations. Sources marked with a `v' in the note column are variable (see Section \ref{sec:var}), and those marked with an `e' are extended in FIRST maps (see Section \ref{sec:morph}). Notes on individual sources marked with a * are given in the Appendix. See Section \ref{sec:results} for details.}
\label{tab:sourcelist}
\begin{center}
\begin{tabular}{l|c|c|r|r|r|r|l|l|}
Name & RA (J2000) & Dec (J2000) & $S_{1.4}$ (mJy) & $S_{4.8}$ (mJy) & $S_{15}$ (mJy) & $S_{30}$ (mJy) & $\alpha_{1.4}^{30}$ & Note\\
\hline
J0011+2830 & 00 11 46.71 & 28 30 23.1 & 128.8 $\pm$ 3.9 & 36 $\pm$ 5 & 8.8 & 4.6 $\pm$ 1.1 & -1.09$^{+0.08}_{-0.10}$\\
J0011+2811 & 00 11 50.57 & 28 11 08.4 & 10.0 $\pm$ 0.5 & -- & 3.6 & -2.6 $\pm$ 1.7 & $<-0.4$ \\
J0011+2757 & 00 11 51.72  & 27 57 18.3 & $<$1.5 & -- & 6.0 & 4.0 $\pm$ 1.0 & -- \\
J0012+2821 & 00 12 01.34 & 28 21 47.1 & 8.4 $\pm$ 0.5 & -- & 6.2 & -0.7 $\pm$ 0.8 & $<-0.5$ \\
J0012+2829 & 00 12 44.44 & 28 29 09.5 & 36.8 $\pm$ 1.2 & -- & 4.4 & 4.6 $\pm$ 1.5 & -0.7 $\pm$ 0.1\\
J0013+2819 & 00 13 28.04 & 28 19 49.5 & 48.6 $\pm$ 1.5 & 23 $\pm$ 4 & 8.4 & 4.2 $\pm$ 1.7 & $<-0.5$\\
J0013+2834 & 00 13 32.68 & 28 34 48.8 & 39.8 $\pm$ 1.3 & 32 $\pm$ 5 & 23.8 & 14.5 $\pm$ 2.0 & -0.33$^{+0.05}_{-0.06}$ & v\\
J0014+2845 & 00 14 14.80 & 28 45 50.9 & 49.4 $\pm$ 2.1 & -- & 4.6 & 1.0 $\pm$ 1.6 & $<-0.7$ \\
J0014+2821 & 00 14 16.75 & 28 21 09.3 & 263.0 $\pm$ 7.9 & 56 $\pm$ 6 & 12.1 & 1.4 $\pm$ 1.9 & $<-1.1$ \\
J0014+2848 & 00 14 24.78 & 28 48 44.3 & 19.7 $\pm$ 0.7 & -- & 2.6 & 0.9 $\pm$ 3.3 & $<-0.2$ \\
(J0014+2852a) & 00 14 27.74 & 28 52 26.8 & 19.1 $\pm$ 1.0 & -- & 5.5 & 2.7 $\pm$ 1.3 & $<-0.3$ & *\\
(J0014+2852b) & 00 14 33.14 & 28 52 05.1 & 6.8 $\pm$ 0.5 & -- & 5.4 & 5.2 $\pm$ 1.0 & -0.09$^{+0.08}_{-0.10}$ & *\\
J0014+2815 & 00 14 33.84 & 28 15 05.3 & 96.0 $\pm$ 2.9 & 135 $\pm$ 12 & 45.1 & 27.6 $\pm$ 2.8 & -0.41 $\pm$ 0.04\\
J0015+2906 & 00 15 13.97 & 29 06 13.0 & 10.6 $\pm$ 0.5 & 20 $\pm$ 4 & 5.6 & 6.5 $\pm$ 1.4 & -0.16$^{+0.08}_{-0.09}$\\
J0015+2831 & 00 15 15.45 & 28 31 53.4 & 3.7 $\pm$ 1.0 &  -- & 3.0 & 5.4 $\pm$ 2.3 & $<1.6$ \\
J0015+2819 & 00 15 36.64 & 28 19 22.9 & 7.1 $\pm$ 0.5 & -- & 2.5 & 1.2 $\pm$ 1.8 & $<0.1$ \\
J0015+2843 & 00 15 42.48 & 28 43 39.8 & 35.7 $\pm$ 1.1 & -- & 7.5 & 6.9 $\pm$ 1.9 & -0.53$^{+0.09}_{-0.11}$\\
J0015+2718 & 00 15 58.02 & 27 18 53.8 & 2.1 $\pm$ 1.5 & -- & 3.1 & -2.9 $\pm$ 1.6 & $<0.4$\\
J0016+2804 & 00 16 02.36 & 28 04 27.4 & 50.2 $\pm$ 1.6 & 20 $\pm$ 4 & 2.7 & 2.1 $\pm$ 2.7 & $<-0.5$ \\
J0017+2733 & 00 17 04.58 & 27 33 42.3 & 11.5 $\pm$ 0.5 & -- & 6.2 & 2.8 $\pm$ 2.0 & $<0.0$ \\
J0017+2750 & 00 17 23.77 & 27 50 27.5 & 302.0 $\pm$ 10.7 & 88 $\pm$ 8 & 20.9 & 11.1 $\pm$ 1.9 & -1.08$^{+0.06}_{-0.07}$\\
J0017+2835 & 00 17 32.50 & 28 35 12.4 & 82.8 $\pm$ 2.5 & 35 $\pm$ 5 & 9.3 & 4.8 $\pm$ 2.1 & $<-0.6$\\
J0017+2736 & 00 17 59.94 & 27 36 16.1 & 72.8 $\pm$ 2.2 & -- & 4.7 & 4.4 $\pm$ 1.2 & -0.92$^{+0.09}_{-0.12}$\\
J0018+2749 & 00 18 09.49 & 27 49 24.5 & 6.2 $\pm$ 0.4 & -- & 10.4 & 7.0 $\pm$ 1.9 & 0.0$^{+0.1}_{-0.1}$\\
J0018+2843 & 00 18 11.99 & 28 43 56.0 & 17.9 $\pm$ 0.7 & -- & 11.1 & 4.8 $\pm$ 1.1 & -0.43$^{+0.08}_{-0.10}$ & v\\
J0018+2846 & 00 18 12.78 & 28 46 37.6 & 153.7 $\pm$ 5.3 & 54 $\pm$ 6 & 15.0 & 9.4 $\pm$ 2.0 & -0.91$^{+0.07}_{-0.09}$\\
J0018+2731 & 00 18 32.72 & 27 31 02.9 & 150.8 $\pm$ 4.5 & 57 $\pm$ 6 & 21.2 & 11.4 $\pm$ 2.0 & -0.84$^{+0.06}_{-0.07}$\\
J0018+2801 & 00 18 42.15 & 28 01 07.9 & 7.5 $\pm$ 1.0 & -- & 3.7 & 1.1 $\pm$ 2.7 & $<0.2$ \\
J0018+2744 & 00 18 58.17 & 27 44 45.5 & 144.3 $\pm$ 4.3 & 42 $\pm$ 5 & 10.0 & 3.3 $\pm$ 1.3 & $<-0.9$ \\
J0019+2817 & 00 19 08.97 & 28 17 54.6 & 24.7 $\pm$ 0.8 & -- & 32.6 & 26.0 $\pm$ 3.5 & 0.02$^{+0.05}_{-0.06}$ & v\\

J0259+2627 & 02 59 55.32 & 26 27 29.2 & 13.8 $\pm$ 0.8 & -- & 9.4 & 5.5 $\pm$ 1.5 & -0.30$^{+0.10}_{-0.12}$\\
J0300+2654 & 03 00 15.58 & 26 54 52.2 & 5.3 $\pm$ 0.4 & -- & 5.4 & 2.7 $\pm$ 1.2 & $<0.1$ \\
J0301+2547 & 03 01 05.43 & 25 47 16.1 & 46.3 $\pm$ 1.8 & 23 $\pm$ 4 & 14.2 & 14.6 $\pm$ 2.2 & -0.38$^{+0.06}_{-0.07}$\\
(J0301+2541) & 03 01 37.19 & 25 41 56.2 & 300.6 $\pm$ 9.8 & 92 $\pm$ 9 & 18.0 & 10.1 $\pm$ 2.4 & -1.11$^{+0.08}_{-0.10}$ & v*\\
J0301+2521 & 03 01 38.18 & 25 21 49.0 & 134.8 $\pm$ 4.1 & 44 $\pm$ 5 & 14.5 & 8.9 $\pm$ 2.0 & -0.89$^{+0.08}_{-0.09}$\\
(J0301+2542) & 03 01 39.80 & 25 42 30.1 & 23.5 $\pm$ 0.8 & 92 $\pm$ 9 & 7.2 & 4.5 $\pm$ 1.5 & -0.5 $\pm$ 0.1 & *\\
J0302+2549 & 03 02 19.94 & 25 49 49.0 & 6.7 $\pm$ 0.4 & -- & 7.3 & 6.9 $\pm$ 1.7 & 0.01$^{+0.09}_{-0.11}$\\
J0302+2607 & 03 02 42.78 & 26 07 53.8 & 77.7 $\pm$ 2.4 & 34 $\pm$ 5 & 11.7 & 6.5 $\pm$ 2.2 & $<-0.5$\\
J0302+2625 & 03 02 35.72 & 26 25 53.2 & 26.6 $\pm$ 0.9 & -- & 3.5 & 2.7 $\pm$ 1.7 & $<-0.4$\\
J0303+2641 & 03 03 31.21 & 26 41 20.9 & 20.9 $\pm$ 1.0 & -- & 5.0 & 3.2 $\pm$ 2.1 & $<-0.2$ \\
J0303+2645 & 03 03 34.79 & 26 45 37.4 & 365.0 $\pm$ 11.0 & 102 $\pm$ 10 & 34.1 & 15.1 $\pm$ 2.1 & -1.04$^{+0.05}_{-0.06}$\\
J0303+2700 & 03 03 58.95 & 27 00 51.0 & 3.0 $\pm$ 0.5 & -- & 6.0 & 3.3 $\pm$ 2.7 & $<0.7$ \\
J0303+2531 & 03 03 59.56 & 25 31 34.2 & 153.9 $\pm$ 4.6 & 122 $\pm$ 11 & 55.1 & 45.6 $\pm$ 5.2 & -0.40$^{+0.04}_{-0.05}$ & v\\
J0304+2659 & 03 04 23.60 & 26 59 11.0 & 6.5 $\pm$ 0.7 &  -- & 4.2 & 1.7 $\pm$ 1.2 & $<0.1$ & *\\
J0304+2640 & 03 04 34.02 & 26 40 03.3 & 52.1 $\pm$ 1.9 & 23 $\pm$ 4 & 6.1 & 5.4 $\pm$ 1.1 & -0.74$^{+0.07}_{-0.09}$\\
J0304+2623 & 03 04 40.82 & 26 23 10.6 & 53.1 $\pm$ 2.0 & -- & 4.6 & 4.1 $\pm$ 1.3 & -0.8 $\pm$ 0.1\\
J0304+2551 & 03 04 58.58 & 25 51 49.5 & 480.8 $\pm$ 14.4 & 116 $\pm$ 11 & 28.9 & 13.4 $\pm$ 1.5 & -1.17$^{+0.04}_{-0.05}$\\
J0305+2532 & 03 05 14.96 & 25 32 16.6 & 20.3 $\pm$ 0.7 & -- & 5.5 & 6.0 $\pm$ 1.3 & -0.40$^{+0.07}_{-0.09}$\\
J0306+2601 & 03 06 25.43 & 26 01 21.5 & 18.7 $\pm$ 0.7 & -- & 7.9 & 12.6 $\pm$ 2.2 & -0.13$^{+0.07}_{-0.08}$\\
J0306+2548 & 03 06 27.66 & 25 48 22.8 & 89.2 $\pm$ 2.7 & 57 $\pm$ 6 & 18.1 & 10.7 $\pm$ 1.7 & -0.69$^{+0.06}_{-0.07}$ & v\\
J0306+2627 & 03 06 32.73 & 26 27 26.3 & 95.2 $\pm$ 2.9 & 25 $\pm$ 4 & 9.1 & 1.9 $\pm$ 2.1 & $<-0.8$\\
J0307+2625 & 03 07 00.33 & 26 25 48.8 & 24.4 $\pm$ 0.8 & 20 $\pm$ 4 & 20.8 & 17.6 $\pm$ 4.1 & -0.11$^{+0.08}_{-0.10}$\\

J0936+3203 & 09 36 37.43 & 32 03 31.6 & 191.2 $\pm$ 5.7 & 73 $\pm$ 7 & 22.2 & 6.1 $\pm$ 1.4 & -1.13$^{+0.08}_{-0.09}$ & v\\
J0936+3118 & 09 36 53.39 & 31 18 25.9 & 21.6 $\pm$ 0.8 & -- & 3.1 & -1.4 $\pm$ 2.8 & $<-0.3$ & e\\
J0936+3129 & 09 36 58.25 & 31 29 29.9 & 75.9 $\pm$ 8.3 &  31 $\pm$ 4 & 5.9 & 3.0 $\pm$ 2.8 & $<-0.5$ & e*\\
J0937+3206 & 09 37 06.42 & 32 06 55.4 & 116.0 $\pm$ 3.5 & 92 $\pm$ 9 & 49.3 & 61.8 $\pm$ 7.9 & -0.21 $\pm$ 0.05 & e\\
J0937+3201 & 09 37 22.63 & 32 01 03.7 & $<$1.5 &  -- & 5.3 & -0.8 $\pm$ 2.2 & -- \\
J0937+3143 & 09 37 58.41 & 31 43 41.2 & 14.6 $\pm$ 0.6 & -- & 6.5 & -2.3 $\pm$ 3.2 & $<-0.2$ \\

\end{tabular}
\end{center}
\end{table*}
\begin{table*}
\contcaption{}
\begin{center}
\begin{tabular}{l|c|c|r|r|r|r|l|l|}
Name & RA (J2000) & Dec (J2000) & $S_{1.4}$ (mJy) & $S_{4.8}$ (mJy) & $S_{15}$ (mJy) & $S_{30}$ (mJy) & $\alpha_{1.4}^{30}$ & Note\\
\hline
J0938+3118 & 09 38 17.79 & 31 18 52.8 & 24.9 $\pm$ 0.9 & 22 $\pm$ 4 & 13.9 & 6.2 $\pm$ 2.2 & $<-0.2$ & e \\
J0939+3134 & 09 39 48.18 & 31 34 01.9 & 34.2 $\pm$ 1.1 & -- & 3.9 & 5.3 $\pm$ 2.0 & $<-0.3$ & e\\
J0939+3154 & 09 39 50.76 & 31 54 13.7 & 96.5 $\pm$ 2.9 & 30 $\pm$ 4 & 9.8 & 6.5 $\pm$ 2.7 & $<-0.6$ \\
J0939+3122 & 09 39 53.86 & 31 22 43.9 & 92.4 $\pm$ 3.2 & 22 $\pm$ 4 & 8.5 & -1.8 $\pm$ 2.5 & $<-0.9$ & e\\
J0940+3240 & 09 40 38.45 & 32 40 04.8 & 30.5 $\pm$ 1.0 & -- & 3.4 & -3.2 $\pm$ 1.4 & $<-1.1$ \\
J0940+3201 & 09 40 41.85 & 32 01 31.2 & 118.5 $\pm$ 3.6 & 34 $\pm$ 5 & 12.3 & 4.6 $\pm$ 1.3 & -1.06$^{+0.09}_{-0.12}$\\
J0941+3126 & 09 41 03.32 & 31 26 20.1 & 111.1 $\pm$ 3.7 & 67 $\pm$ 7 & 20.0 & 7.2 $\pm$ 3.3 & $<-0.6$ & e*\\
J0941+3221 & 09 41 05.86 & 32 21 45.3 & 381.0 $\pm$ 11.4 & 93 $\pm$ 9 & 20.2 & 11.1 $\pm$ 1.8 & -1.16$^{+0.06}_{-0.07}$\\
J0941+3057 & 09 41 16.03 & 30 57 29.2 & 61.2 $\pm$ 2.3 & 22 $\pm$ 4 & 4.4 & -1.0 $\pm$ 2.3 & $<-0.7$ & e\\
J0941+3146 & 09 41 46.95 & 31 46 47.9 & 3.3 $\pm$ 0.4 & -- & 3.1 & 0.8 $\pm$ 1.3 & $<0.3$ \\
J0941+3154 & 09 41 47.06 & 31 54 53.2 & 288.5 $\pm$ 9.4 & 99 $\pm$ 9 & 22.7 & 12.7 $\pm$ 2.3 & -1.02$^{+0.07}_{-0.08}$ & e\\
J0941+3226 & 09 41 47.47 & 32 26 46.1 & 76.1 $\pm$ 1.2 &  36 $\pm$ 5 & 5.4 & 2.2 $\pm$ 1.8 & $<-0.7$ & e*\\
J0942+3239 & 09 42 00.71 & 32 39 04.5 & 16.4 $\pm$ 0.6 & -- & 17.9 & 8.2 $\pm$ 3.2 & $<0.1$ \\
J0942+3206 & 09 42 08.09 & 32 06 40.7 & 189.3 $\pm$ 6.7 & 62 $\pm$ 7 & 21.4 & 10.5 $\pm$ 2.4 & -0.94$^{+0.08}_{-0.10}$ & e\\
J0942+3150 & 09 42 54.23 & 31 50 50.6 & 105.2 $\pm$ 3.9 & 37 $\pm$ 5 & 11.8 & 6.4 $\pm$ 2.6 & $<-0.6$ & e\\
J0943+3159 & 09 43 08.16 & 31 59 38.0 & 30.9 $\pm$ 1.0 & -- & 12.5 & 2.2 $\pm$ 2.4 & $<-0.4$ & ve \\
J0943+3132 & 09 43 44.12 & 31 32 07.1 & 33.3 $\pm$ 1.4 & -- & 3.4 & -1.3 $\pm$ 2.4 & $<-0.5$ & e\\
J0944+3115 & 09 44 11.60 & 31 15 21.0 & 56.2 $\pm$ 2.1 & 31 $\pm$ 4 & 24.3 & 32.1 $\pm$ 3.1 & -0.18 $\pm$ 0.04 & e\\

J1227+5240 & 12 27 17.93 & 52 40 02.9 & 26.8 $\pm$ 0.9 & -- & 2.6 & 2.4 $\pm$ 1.5 & $<-0.4$ & e\\
J1227+5313 & 12 27 52.75 & 53 13 45.7 & 12.8 $\pm$ 0.5 & -- & 5.5 & 4.7 $\pm$ 1.5 & -0.3 $\pm$ 0.1\\
J1228+5308 & 12 28 19.26 & 53 08 24.6 & 6.5 $\pm$ 0.4 & -- & 12.0 & 2.8 $\pm$ 1.3 & $<0.1$ \\
J1230+5230 & 12 30 08.85 & 52 30 51.6 & 74.8 $\pm$ 2.3 & 24 $\pm$ 4 & 7.5 & -1.0 $\pm$ 1.3 & $<-1.0$ & e\\
J1232+5202 & 12 32 18.49 & 52 02 19.2 & 93.0 $\pm$ 2.8 & 21 $\pm$ 4 & 9.2 & 3.9 $\pm$ 2.3 & $<-0.7$ & e\\
J1233+5339 & 12 33 11.08 & 53 39 56.9 & 62.6 $\pm$ 1.9 & 26 $\pm$ 5 & 24.9 & 21.3 $\pm$ 2.5 & -0.35 $\pm$ 0.05\\
J1233+5337 & 12 33 41.38 & 53 37 27.7 & 80.8 $\pm$ 2.5 & -- & 6.7 & 3.6 $\pm$ 0.8 & -1.01$^{+0.08}_{-0.10}$ &\\
J1234+5322 & 12 34 35.89 & 53 22 35.7 & 44.0 $\pm$ 1.7 & 24 $\pm$ 4 & 5.2 & 4.4 $\pm$ 1.2 & -0.75$^{+0.09}_{-0.11}$ & e\\
J1235+5317 & 12 35 00.71 & 53 17 58.9 & 5.7 $\pm$ 1.0  &  -- & 4.0 & 4.5 $\pm$ 1.0 & -0.1 $\pm$ 0.1 & *\\
J1235+5311 & 12 35 22.86 & 53 11 29.0 & $<$1.5  &  -- & 6.2 & 5.4 $\pm$ 2.1 & -- & *\\
J1235+5210 & 12 35 29.92 & 52 10 01.4 & 166.0 $\pm$ 5.0 & 41 $\pm$ 5 & 10.5 & 3.7 $\pm$ 1.3 & $<-1.0$ \\
J1235+5228 & 12 35 30.53 & 52 28 27.3 & 87.5 $\pm$ 2.7 & 80 $\pm$ 8 & 47.0 & 27.3 $\pm$ 2.6 & -0.38 $\pm$ 0.04\\
J1235+5315 & 12 35 27.86 & 53 15 01.7 & 49.2 $\pm$ 2.1 & -- & 6.6 & 2.6 $\pm$ 1.4 & $<-0.6$ & e\\
J1237+5325 & 12 37 02.19 & 53 25 25.1 & 71.5 $\pm$ 2.6 & -- & 9.0 & 5.1 $\pm$ 2.8 & $<-0.5$ & e\\
J1237+5254 & 12 37 04.16 & 52 54 22.4 & 373.5 $\pm$ 14.4 & 126 $\pm$ 11 & 35.4 & 20.1 $\pm$ 2.2 & -0.95 $\pm$ 0.05 & e\\
J1237+5303 & 12 37 25.13 & 53 03 14.9 & 51.9 $\pm$ 1.9 & 24 $\pm$ 4 & 8.1 & 3.9 $\pm$ 1.3 & -0.8 $\pm$ 0.1& e\\
(J1238+5249a) & 12 38 32.28 & 52 49 15.8 & 93.1 $\pm$ 2.8 & -- & 8.3 & 3.7 $\pm$ 1.1 & -1.05$^{+0.09}_{-0.12}$ & *\\
(J1238+5249b) & 12 38 49.78 & 52 49 34.3 & 324.9 $\pm$ 9.8 & 103 $\pm$ 10 & 34.2 & 17.5 $\pm$ 2.1 & -0.95 $\pm$ 0.05 & e*\\

J1535+4103 & 15 35 00.62 & 41 03 13.6 & 79.1 $\pm$ 2.4 & 19 $\pm$ 4 & 5.9 & -1.3 $\pm$ 2.2 & $<-0.8$ & e\\
(J1535+4143a) & 15 35 02.46 & 41 43 04.8 & 51.8 $\pm$ 2.2 & 67 $\pm$ 7 & 3.5 & 7.0 $\pm$ 1.2 & -0.65$^{+0.06}_{-0.07}$ & e*\\
(J1535+4143b) & 15 35 05.68 & 41 43 28.2 & 100.8 $\pm$ 3.7 & as above & 2.5 & 6.4 $\pm$ 1.9 & -0.90$^{+0.10}_{-0.13}$ & e*\\
(J1535+4142) & 15 35 14.72 & 41 42 54.0 & 85.2 $\pm$ 2.6 & as above & 4.0 & 5.2 $\pm$ 2.5 & $<-0.6$ & e*\\
J1536+4125 & 15 36 18.07 & 41 25 38.5 & 78.5 $\pm$ 2.4 & 36 $\pm$ 5 & 8.9 & 3.9 $\pm$ 2.4 & $<-0.6$\\
J1536+4219 & 15 36 49.85 & 42 19 57.5 & 60.5 $\pm$ 2.2 & -- & 6.3 & 2.4 $\pm$ 1.5 & $<-0.7$ & e\\
J1537+4104 & 15 37 11.91 & 41 04 13.0 & 4.8 $\pm$ 1.1 &  -- & 3.3 & 1.0 $\pm$ 1.5 & $<0.4$ & e\\
J1537+4212 & 15 37 14.86 & 42 12 53.6 & 16.4 $\pm$ 0.9 & -- & 9.8 & 10.2 $\pm$ 2.4 & -0.16$^{+0.09}_{-0.10}$\\
J1538+4114 & 15 38 13.93 & 41 14 06.9 & 12.4 $\pm$ 0.5 & -- & 7.5 & 2.1 $\pm$ 2.1 & $<-0.1$ & e\\
J1538+4158 & 15 38 15.57 & 41 58 28.4 & 29.9 $\pm$ 1.0 & -- & 5.8 & 0.8 $\pm$ 2.2 & $<-0.4$ \\
J1538+4105 & 15 38 18.59 & 41 05 48.4 & 60.7 $\pm$ 1.9 & 33 $\pm$ 5 & 25.4 & 27.6 $\pm$ 2.5 & -0.26 $\pm$ 0.04 & e\\
J1538+4044 & 15 38 32.10 & 40 44 55.3 & 31.3 $\pm$ 1.0 & 20 $\pm$ 4 & 3.2 & 4.4 $\pm$ 1.6 & $<-0.4$ & e\\
J1538+4215 & 15 38 47.13 & 42 15 27.1 & 230.5 $\pm$ 8.0 & 89 $\pm$ 8 & 25.6 & 14.9 $\pm$ 2.2 & -0.89 $\pm$ 0.06 & e\\
J1538+4225 & 15 38 55.81 & 42 25 27.0 & 57.4 $\pm$ 2.1 & 44 $\pm$ 5 & 31.9 & 27.5 $\pm$ 3.4 & -0.24 $\pm$ 0.05 & ve\\
J1539+4120 & 15 39 10.52 & 41 20 13.3 & 4.3 $\pm$ 0.4 & -- & 4.0 & 6.3 $\pm$ 2.0 & 0.1$^{+0.1}_{-0.2}$ & e\\
J1539+4217 & 15 39 25.63 & 42 17 28.3 & 49.0 $\pm$ 1.5 & 29 $\pm$ 4 & 31.7 & 32.1 $\pm$ 3.0 & -0.14 $\pm$ 0.04\\
J1539+4123 & 15 39 36.78 & 41 23 34.8 & 13.8 $\pm$ 0.6 & 26 $\pm$ 4 & 31.8 & 36.2 $\pm$ 3.3 & 0.31 $\pm$ 0.04\\
J1539+4148 & 15 39 38.97 & 41 48 51.8 & 3.5 $\pm$ 0.4 & -- & 3.2 & 0.3 $\pm$ 1.2 & $<0.2$ & e\\
J1539+4143 & 15 39 51.38 & 41 43 26.4 & 23.2 $\pm$ 0.8 & -- & 17.3 & 14.9 $\pm$ 2.7 & -0.14$^{+0.07}_{-0.08}$\\
J1540+4221 & 15 40 42.21 & 42 21 37.3 & 62.0 $\pm$ 2.5 & 24 $\pm$ 4 & 2.5 & 2.2 $\pm$ 1.1 & $<-0.7$ & e\\
J1540+4138 & 15 40 43.09 & 41 38 17.4 & 19.4 $\pm$ 0.7 & 20 $\pm$ 4 & 45.6 & 25.3 $\pm$ 3.3 & 0.09$^{+0.05}_{-0.06}$ & v\\
J1541+4114 & 15 41 01.24 & 41 14 27.5 & 48.3 $\pm$ 0.8 &  53 $\pm$ 6 & 37.8 & 15.9 $\pm$ 2.3 & -0.36$^{+0.05}_{-0.06}$ & ve*\\
J1541+4049 & 15 41 07.17 & 40 49 17.2 & 159.7 $\pm$ 5.3 & 37 $\pm$ 5 & 10.0 & 4.4 $\pm$ 1.0 & -1.17$^{+0.08}_{-0.10}$ & e\\
J1541+4122 & 15 41 47.41 & 41 22 24.2 & 7.1 $\pm$ 0.5 & -- & 3.5 & 1.4 $\pm$ 1.5 & $<0.0$ & e\\
J1542+4147 & 15 42 40.82 & 41 47 48.3 & 3.2 $\pm$ 0.4 & -- & 5.2 & 4.8 $\pm$ 1.3 & 0.1 $\pm$ 0.1\\
J1543+4152 & 15 43 14.37 & 41 52 24.0 & 4.2 $\pm$ 0.5 & -- & 5.6 & 2.0 $\pm$ 1.2 & $<0.2$\\
J1543+4139 & 15 43 33.09 & 41 39 32.7 & 37.9 $\pm$ 1.2 & -- & 5.6 & 2.0 $\pm$ 1.2 & $<0.2$ \\
\end{tabular}
\end{center}
\end{table*}

The source names and J2000.0 positions at 15~GHz are listed in Table \ref{tab:sourcelist}, along with their flux densities at 15~GHz from the RT observations and at 30~GHz from the OCRA-p observations. We find flux densities for the sources at 1.4 and 4.8~GHz from the NVSS \citep{Condon1998}\footnote{Accessed via \url{http://www.cv.nrao.edu/nvss/}} and GB6 \citep{Gregory1996}\footnote{Accessed via \url{ http://heasarc.gsfc.nasa.gov/W3Browse/all/gb6.html}} catalogues by searching for sources within 15 and 60~arcsec, respectively, of the 15~GHz source positions. Flux densities for all but a few sources were found in NVSS, although 50 sources were not present in GB6 principally due to the higher flux density limit of this survey. We calculate the two-point 1.4 to 30~GHz spectral index for the sources, defined such that $S_\nu \propto \nu^\alpha$.

\subsection{Distribution of flux densities}
\begin{figure}
\centering
\includegraphics[width=7cm]{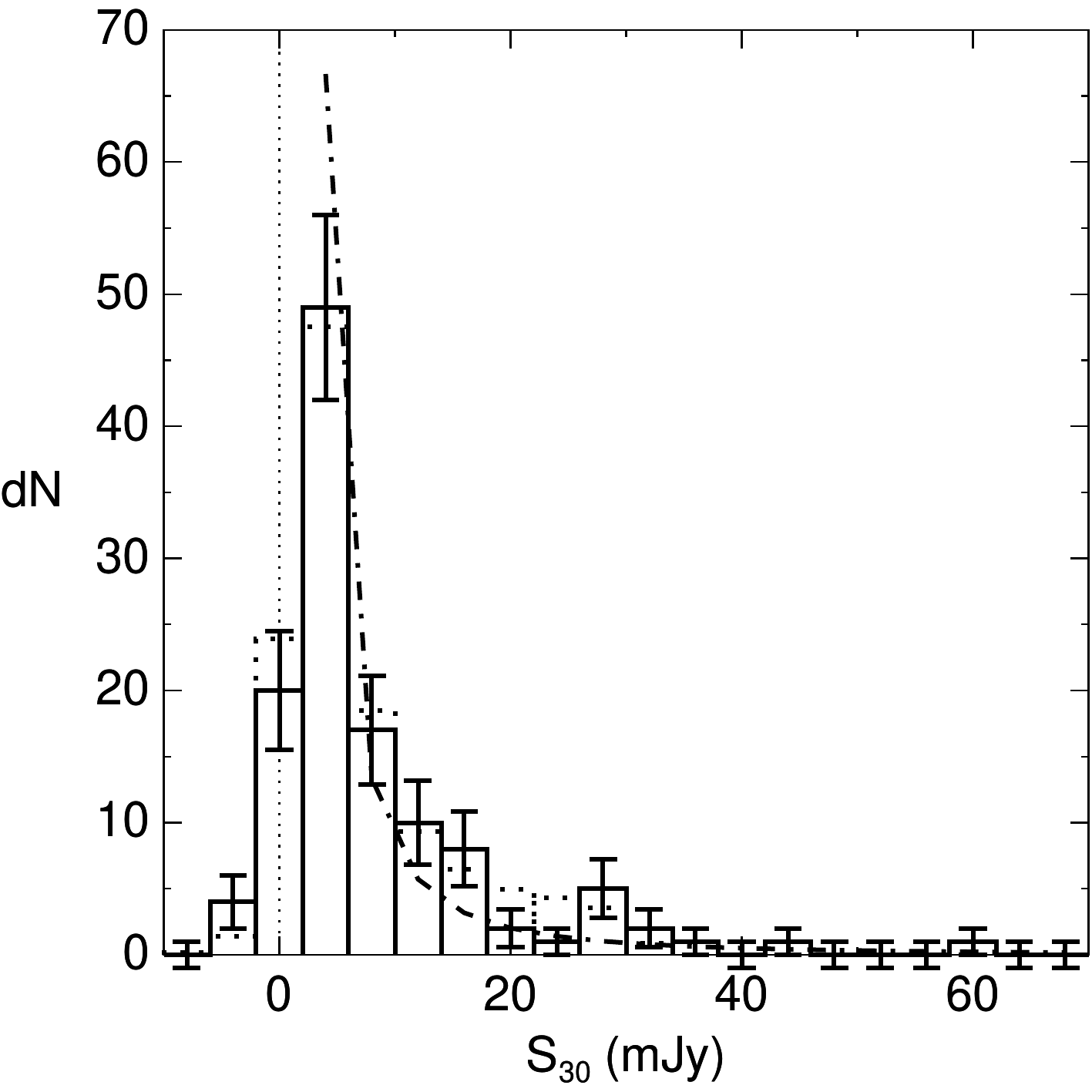}
\caption{The binned numbers of the source flux densities at 30~GHz, including Poisson errors  $(\sqrt{\max(\mathrm{d}N,1)})$ (solid histogram). The simple power law model is also shown (dot-dashed curve). Additionally, a simulated observational histogram (dotted histogram) shows how negative flux densities can occur due to Gaussian errors.}
\label{fig:singlepowerlaw}
\end{figure}

We plot the distribution of 30~GHz flux densities in Figure \ref{fig:singlepowerlaw}. The negative flux densities for several of the sources listed in Table~\ref{tab:sourcelist} are consistent with our error estimates. This can be illustrated using a simple model of the set of observed sources. We assume that the sources have been selected randomly from an intrinsic single power law model described by $N (S_{30}) = 200 \;(S_{30}/1 \mathrm{mJy})^{-1}$ over the whole survey area, with a sharp minimum flux density limit of $S_{30} = 1.7$~mJy, which is the median observational uncertainty. Figure~\ref{fig:singlepowerlaw} also shows the model and the average histogram from 100 simulated sets of values that would be observed if the model is correct and if the errors are Gaussian errors as listed in Table~\ref{tab:sourcelist}. The histogram of our observations shows reasonable consistency with the model, i.e. the negative flux densities are consistent with random Gaussian errors.

Note that the 1.7~mJy flux limit used here is well below the completeness limit of about 10~mJy (see Section \ref{sec:completeness}), so this power law model only provides a rough description of the sample {\em including} selection effects; it does not represent the intrinsic source count distribution.

\subsection{Source confusion}
It is possible that some of the `off' measurements fortuitously fell on previously uncatalogued 30~GHz sources rather than background sky. An upper limit to the level of confusion can be made as follows, by assuming that our sample is complete down to the median observational uncertainty of 1.7~mJy. Since the real completeness limit is about 10~mJy, this should provide a conservative upper limit. Based on the estimate of 31~GHz source counts from \cite{Mason2009}, $N(S_{31}) \sim 17~\mathrm{deg}^{-2} (S_{31}/1 \mathrm{mJy})^{-0.8}$, we expect about 0.003 sources/arcmin$^2$ above 1.7~mJy. A compatible value is obtained from the much smaller 30~GHz survey by \citet{Coble2007}.  The OCRA-p beam width is 1.2~arcmin, so the chance of a single `off' measurement landing on an uncatalogued astrophysical object is about 0.003.  For any single source, the flux density is estimated as an average from 3--26 (mean: 8.3) individual measurements, each of which uses half the signal from each of two opposite `off' positions (see section \ref{sec:datareduction}). The individual measurements for a source are generally made at different position angles, with 3.1~arcmin between `on' and `off' positions. We have 121 sources with 1009 observations of  `off' positions. Hence, the expected number of `off' flux densities which are contaminated by uncatalogued objects is about 3.5.  The chance that more than one of four contaminating `off' flux densities occur for the same source is $\approx 4.8$ per cent,\footnote{The calculation is  equivalent to that of the Birthday Problem: $p = 1 - [(1-1/121) \times (1-2/121) \times (1-3/121)] \approx 0.049$.} assuming that clustering effects are negligible.  Hence, it would be statistically expected that about four sources each have one contaminated `off' flux density measurement, and no sources have multiple contaminations.

For any of the sources affected by one contaminated `off' flux density measurement, the averaging procedure for the multiple measurements is likely to make the effect of the contamination small, since on average 8.3 measurements are made, and a single `off' flux only contributes half of its value to a given measurement.  Hence, on statistical grounds we expect that confusion is negligible for our sample as a whole, though it is likely to have a small effect on a few percent of our sources.

In practice we have direct knowledge of the confusing fields from the RT 15~GHz observations themselves and from the NVSS \citep{Condon1998} and FIRST \citep{White1997} catalogues at 1.4~GHz. In a few instances, sources in the RT 15~GHz list are either two parts of the same source or unrelated sources sufficiently close to each other in the sky as to be confused in the present OCRA observations; these are identified by brackets in Table \ref{tab:sourcelist}. We have examined the NVSS images of all sources and, where available, FIRST images. In the following cases, we conclude that our measured 30~GHz flux densities should be treated with circumspection: J0014+2852a/b, J1535+4142/1535+4143a/b, J0301+2541/42 and J1238+5249a/b. More details on these particular sources are given in the Appendix.

\section{Discussion} \label{sec:discussion}

\subsection{Comparison with previous measurements and source variability} \label{sec:var}
\begin{table*}
\begin{center}
\caption{Comparison of the 15 and 30~GHz flux densities from the RT and OCRA-p (in mJy) from common sources measured with the VLA at 1.4, 4.8, 15.2, 22 and 43~GHz by \citet{Bolton2004,Waldram2007} and those measured by the source subtractor \citep{Cleary2005,Cleary2008}.}
\label{tab:compare_to_cleary}
\begin{tabular}{cccccccccc}
& \multicolumn{5}{c}{VLA} & RT & OCRA & VSA\\
\hline
Name & $S_{1.4}$ & $S_{4.8}$ & $S_{15.2}$ & $S_{22}$ & $S_{43}$ & $S_{15}$ & $S_{30}$ & $S_{33}$ & Notes\\
\hline
J0013+2834 & 32.7 & 33.1 & 34.6 & 36.6 & 30.5 & 23.8 $\pm$ 1.2 & 14.5 $\pm$ 2.0 & 36.8 $\pm$ 1.8 & Variable\\
J0014+2815 & 80.4 & 60.1 & 45.5 & 37.7 & 23.8 & 45.1 $\pm$ 2.3 & 27.6 $\pm$ 2.8 & 34.8 $\pm$ 2.6 &\\
J0019+2817 & 25.9 & 23.8 & 17.4 & 27.0 & 34.9 & 32.6 $\pm$ 1.6 & 26.0 $\pm$ 3.5 & 23.9 $\pm$ 1.7 &  Variable (Fig \ref{fig:0016})\\
J0303+2531 & -- & -- & -- & -- & -- & 55.1 $\pm$ 2.8 & 45.6 $\pm$ 5.2 & 30.0 $\pm$ 2.1 &\\
J0937+3206 & 108.9 & 53.9 & 58.4 & 58.8 & 41.0 & 49.3 $\pm$ 2.5 & 61.8 $\pm$ 7.9 & --\\
J0944+3115 & -- & -- & -- & -- & -- & 24.3 $\pm$ 1.2 & 32.1 $\pm$ 3.1 & 25.3 $\pm$ 2.6 &\\
J1235+5228 & -- & -- & -- & -- & -- & 47.0 $\pm$ 2.4 & 27.3 $\pm$ 2.6 & 27.9 $\pm$ 1.8 &\\
J1539+4123 & -- & -- & -- & -- & -- & 31.8 $\pm$ 1.6 & 36.2 $\pm$ 3.3 & 33.7 $\pm$ 2.4 &\\
J1539+4217 & 53.3 & 40.0 & 34.2 & 37.0 & 26.3 & 31.7 $\pm$ 1.6 & 32.1 $\pm$ 3.0 & 34.0 $\pm$ 1.6 &\\
J1538+4225 & 42.0 & 40.5 & 41.6 & 42.2 & 29.0 & 31.9 $\pm$ 1.6 & 27.5 $\pm$ 3.4 & 41.0 $\pm$ 1.5 & Variable\\
J1540+4138 & 16.0 & 30.9 & 34.2 & 23.0 & 9.1 & 45.6 $\pm$ 2.3 & 25.3 $\pm$ 3.3 & -- &\\
J1541+4114 & 65.0 & 38.0 & 30.5 & 27.5 & 19.2 & 37.8 $\pm$ 1.9 & 15.9 $\pm$ 2.3 & 32.5 $\pm$ 2.3 & Variable\\
\end{tabular}
\end{center}
\end{table*}

The stronger sources ($> 10$mJy at 15~GHz) within the 33 extended VSA fields were observed during the initial VSA campaign at 33~GHz using the VSA source subtractor, with those greater than 20~mJy at 33~GHz used for direct subtraction from the CMB data \citep{Dickinson2004}. A subsample of these was published by \citet{Cleary2005,Cleary2008}. Table \ref{tab:compare_to_cleary} lists the flux densities measured by the source subtractor for the sources in \citet{Cleary2005, Cleary2008} that also lie within the super-extended array fields, as well as measurements of those sources made at a range of frequencies by \citet{Bolton2004} and \citet{Waldram2007} using the Very Large Array, and compares them to the measurements by the RT and OCRA. A number of the sources show evidence of variability.

\begin{figure}
\centering
\includegraphics[scale=0.34]{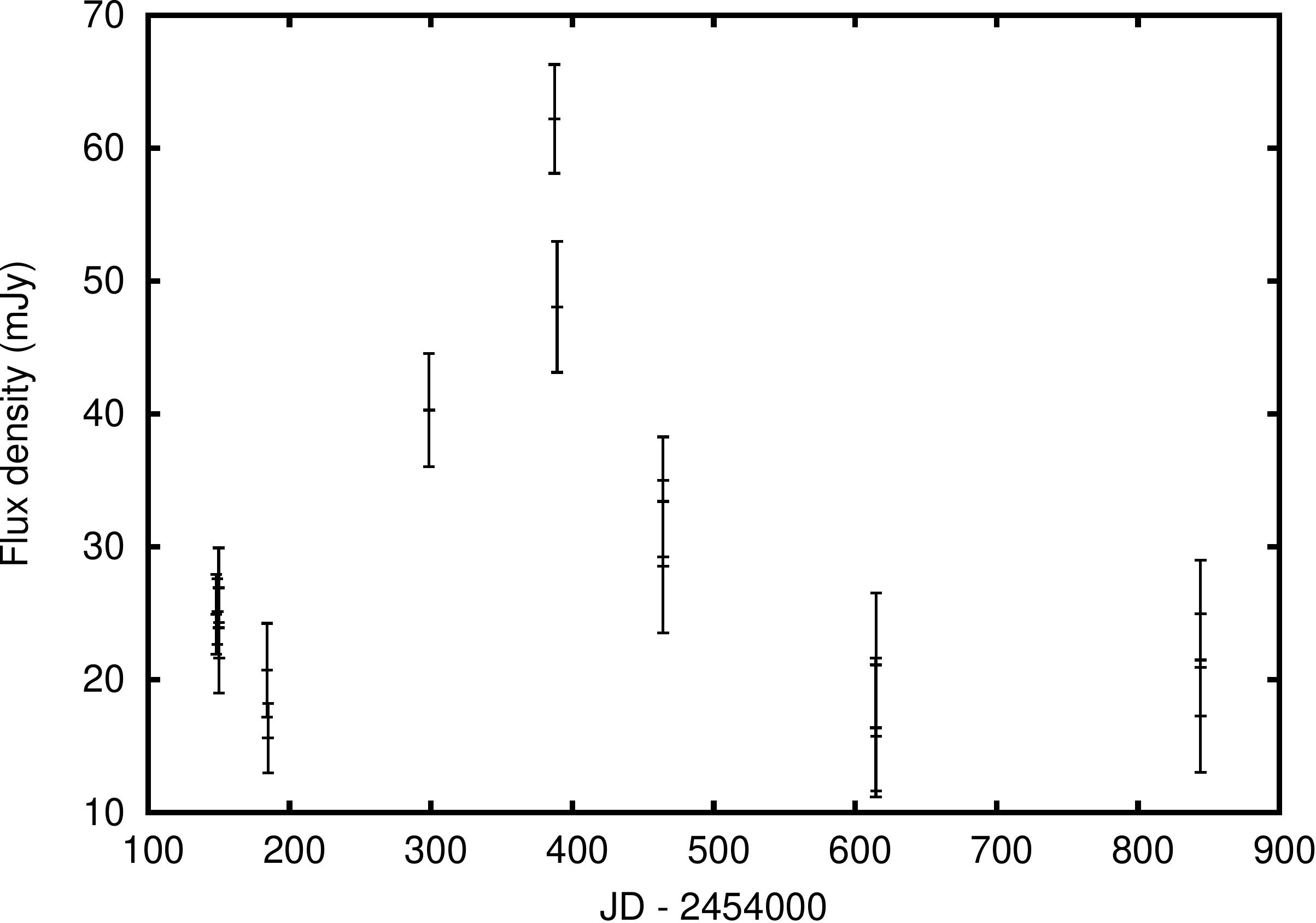}
\caption{The flux density measured for J0019+2817 as a function of time. The source is clearly variable.}
\label{fig:0016}
\end{figure}

We note that OCRA-p is only sensitive to a single linear polarization, which could result in discrepancies between measurements from the VSA, VLA and OCRA for polarized sources; however these are likely to be small since the WMAP sources selected at 22~GHz (see Jackson et al. in preparation) exhibit average linear polarizations of $\sim$3 per cent.  The sources observed with the higher resolution interferometers (RT and VLA) may also be slightly resolved. Nevertheless we identify at least four sources (J0013+2834; J1538+4225; J1541+4114; J0019+2817) as likely to be intrinsically variable. One of these sources -- J0019+2817 -- shows clear variability from OCRA data alone (see Figure \ref{fig:0016}), although the mean flux density from the measurements agrees well with the source subtractor measurement.

\begin{figure}
\centering
\includegraphics[scale=0.34]{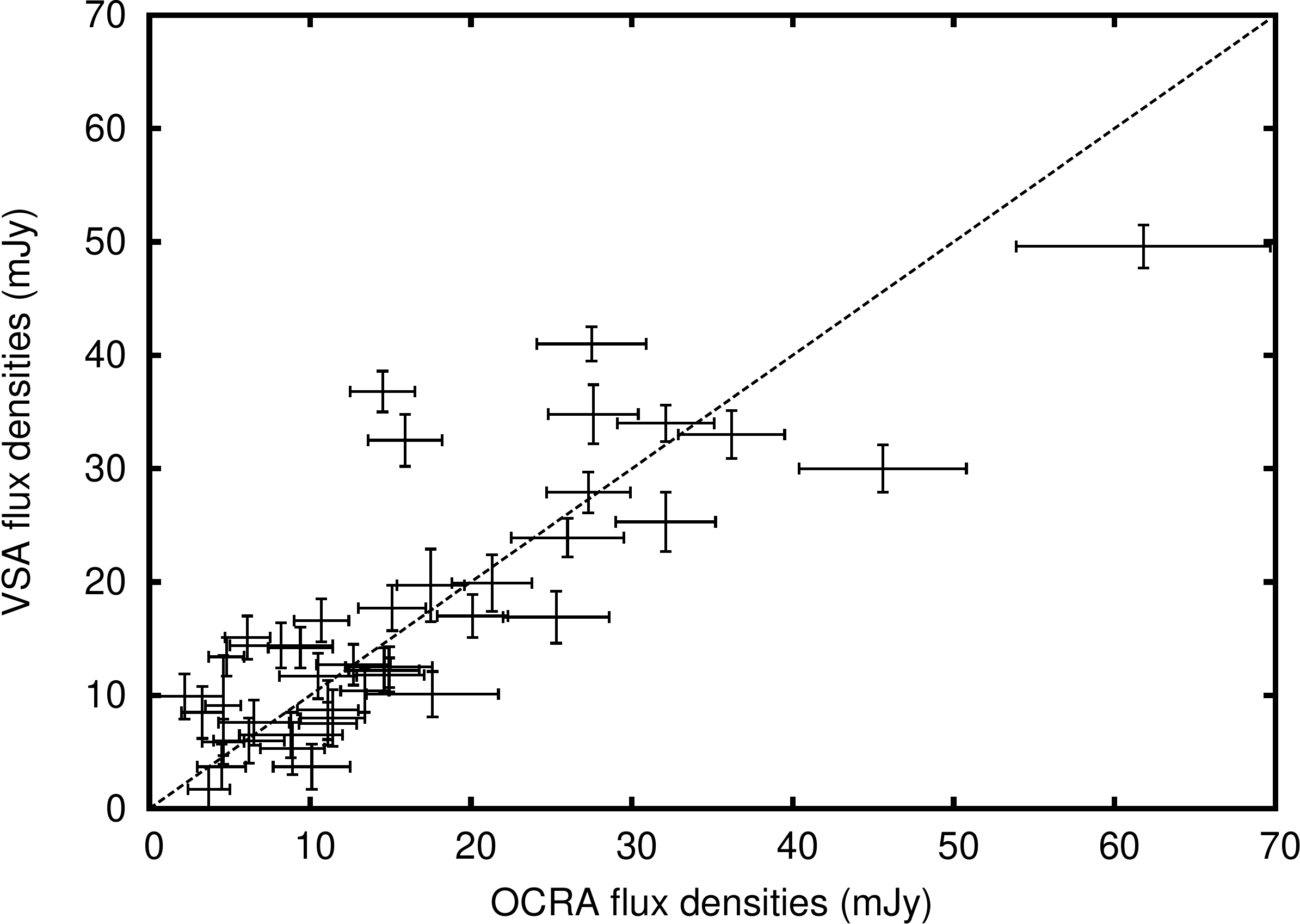}
\caption{Comparison of the OCRA-p flux densities against those from the VSA source subtractor.}
\label{fig:flux_comparison}
\end{figure}

A comparison of flux densities for 42 sources observed both with the VSA source subtractor and OCRA-p is shown in Figure \ref{fig:flux_comparison}. The flux density scales are consistent, however 10 out of the 42 sources show variations greater than $2 \sigma$ (combined error), some of which show variations greater than a factor of 2. These are denoted with a `v' in Table~\ref{tab:sourcelist}.

\subsection{Source spectra} \label{sec:spectra}
\begin{figure}
\centering
\includegraphics[scale=0.34]{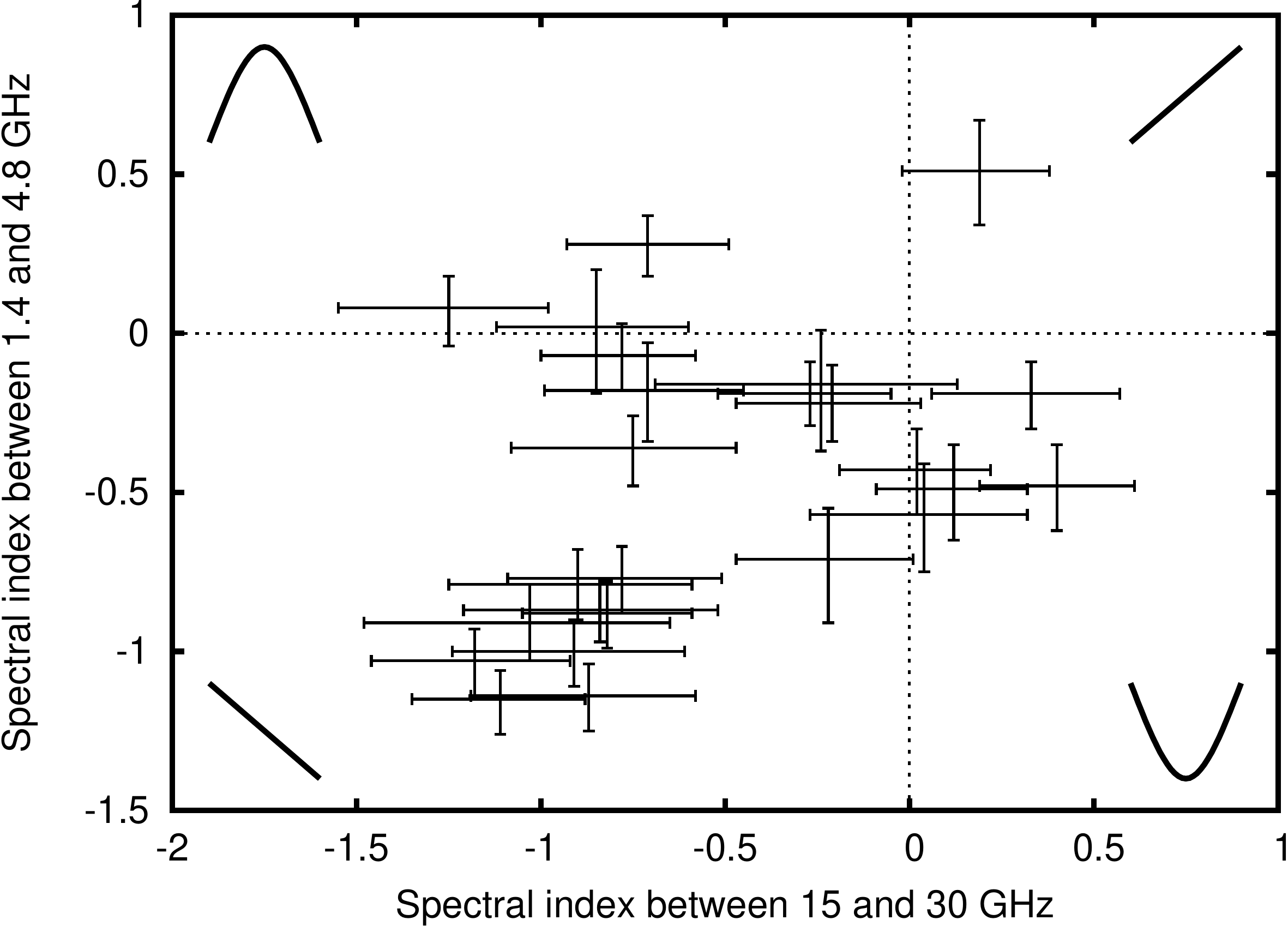}
\caption{$\alpha_{1.4}^{4.8}$ vs. $\alpha_{15}^{30}$ for all sources in this sample with a 30~GHz flux density greater than 10~mJy, and known flux densities at all four frequencies. The diagrams in the corners schematically illustrate the spectral behaviour of the sources in each quadrant.}
\label{fig:2colour}
\end{figure}

Figure \ref{fig:2colour} shows the `2-colour' diagram for sources detected at all four frequencies (1.4, 4.8, 15, 30~GHz). We find few sources with peaked spectra (top left quadrant of Figure \ref{fig:2colour}), but a larger number of sources with spectra rising towards high frequencies (right hand quadrants of Figure \ref{fig:2colour}).

Figure \ref{fig:1p4to30} displays the distribution of spectral indices $S \propto \nu^\alpha$ between 1.4 and 30~GHz for 60 sources above the completeness limit of 7~mJy at 15~GHz (5 out of a total of 65 sources are excluded due to their complex morphology -- 0258+2530a/b, 0938+3140 and 1236+5305/6). There is clear evidence of bimodality, with a split around a spectral index of -0.5, dividing the sources into groups of sources with `flat' and `steep' spectra.

The distribution is markedly different from the 1.4 to 30~GHz spectral index distribution presented by \citet{Mason2009}. However, their source sample was selected at 1.4~GHz rather than 15~GHz as in this paper. The distribution in \citet{Mason2009} is not bimodal, but has a single peak around $-1.0$, implying that their sample is dominated by the steep spectrum sources. Our sample contains a larger fraction of flat and rising spectrum sources. This is not surprising given the difference of a factor ten in the sample selection frequency.

We return to the issue of spectral indices in different samples in sections 6.3 and 6.4.

\subsection{The 1.4-30~GHz spectral index distribution vs flux density }
We are seeking to improve our knowledge of the source population at 30~GHz down to mJy flux densities and beyond (by extrapolation). In order to extrapolate one needs to know if there is a dependence of the spectral properties of the population on flux density at 30~GHz. We have therefore used the WMAP 5-year point source catalogue \citep{Wright2009} to select a complete strong source sample consisting of all sources with a 22~GHz flux above 1~Jy. We have then cross-identified these with the NVSS catalogue and calculated the spectral index distribution between 1.4 and 33~GHz. We chose 22 GHz as the selection frequency to be as close as possible to the 15~GHz selection frequency used in the present paper.

\begin{figure}
\centering
\includegraphics[scale=0.34]{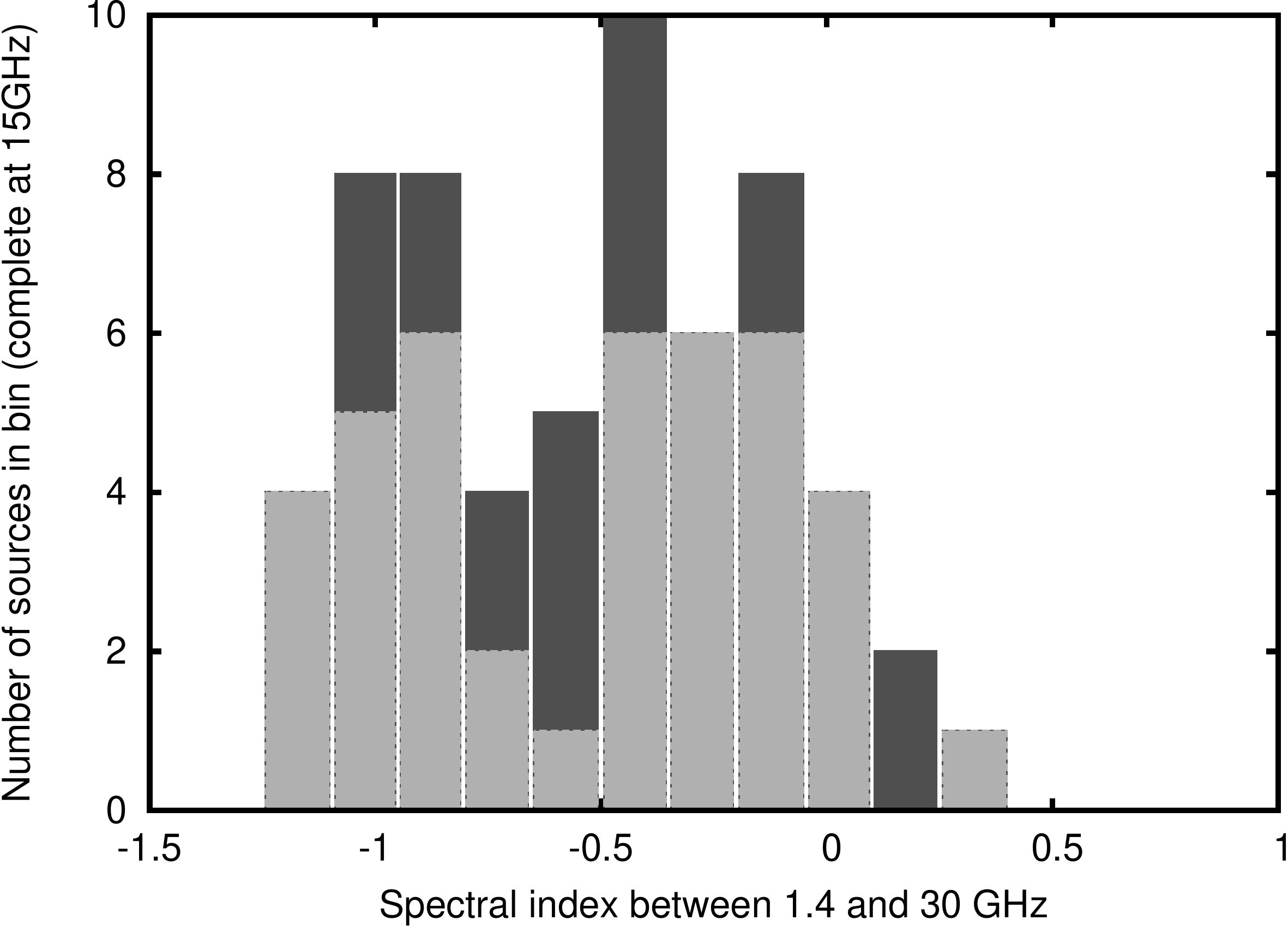}
\caption{The distribution of spectral indices between 1.4 and 30~GHz ($\alpha_{1.4}^{30}$) for 60 sources in this sample with a 15~GHz flux density greater than 7~mJy. $3 \sigma$ detections at 30~GHz are shown in the light grey histogram; the dark grey histogram contains the $3 \sigma$ upper limits on the spectral indices.}
\label{fig:1p4to30}
\end{figure}

\begin{figure}
\centering
\includegraphics[scale=0.34]{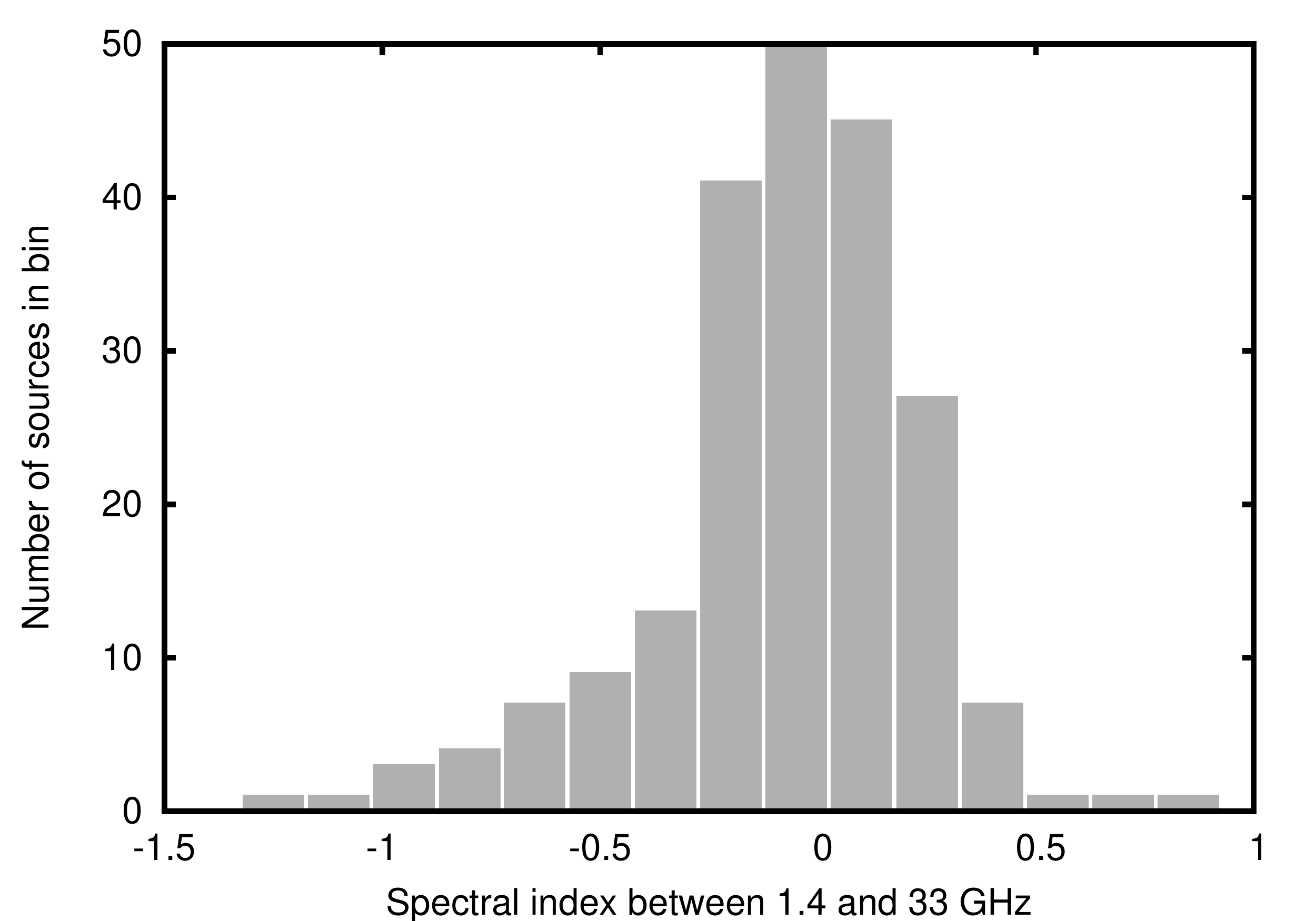}
\caption{The spectral index distribution of 22~GHz-selected WMAP sources between 1.4 and 33~GHz. The majority of sources have flat spectra ($\alpha > -0.5$).}
\label{fig:wmap_indices}
\end{figure}

In this analysis, we exclude sources where there are no obvious NVSS matches, sources known to be extended on scales greater than $\sim30$~arcmin (WMAP J0322-3711/Fornax A; WMAP J1633+8226/NGC~6251) or confused in WMAP (WMAP J0223+4303/3C66 A and B). Sources are matched to single NVSS sources, except for WMAP J0108+1319/3C33 where four close NVSS sources are combined. In total, 211 sources are used in the analysis. Figure \ref{fig:wmap_indices} shows the spectral index distribution, which has a single peak centered on zero. All but 16 sources are classified as flat spectrum, i.e. have $\alpha > -0.5$. As expected, this is also very different to the distribution found by \citet{Mason2009}, which is dominated by steep spectrum sources.

There is no evidence for a significant population of steep spectrum sources in the WMAP spectral index distribution. This contrasts with the VSA source distribution shown in Figure \ref{fig:1p4to30} where there is evidence for both flat and steep spectrum populations. This difference could be ascribed to the selection frequencies (22~GHz cf. 15~GHz), however their closeness suggests that this should not be a major issue. Another consideration is that although we are selecting using the 22~GHz flux densities, in reality the WMAP point source catalogue is obtained using all of the bands \citep{Wright2009}. This should also not be an issue as the sources with 22~GHz flux density greater than 1~Jy are detected by WMAP at high significance. A final possibility, which we think is most likely, is that the dissimilar distributions are due to the difference in the flux densities of the sources in the samples. The population of steep spectrum sources increases as the flux densities decrease.

Such an effect is also seen in \citet{Waldram2009}, where they find a significant change in spectral index distribution with flux density amongst their 15~GHz selected sources. We note, however, that there is some overlap in sources between their samples and the one studied here.

\begin{figure}
\centering
\includegraphics[scale=0.34]{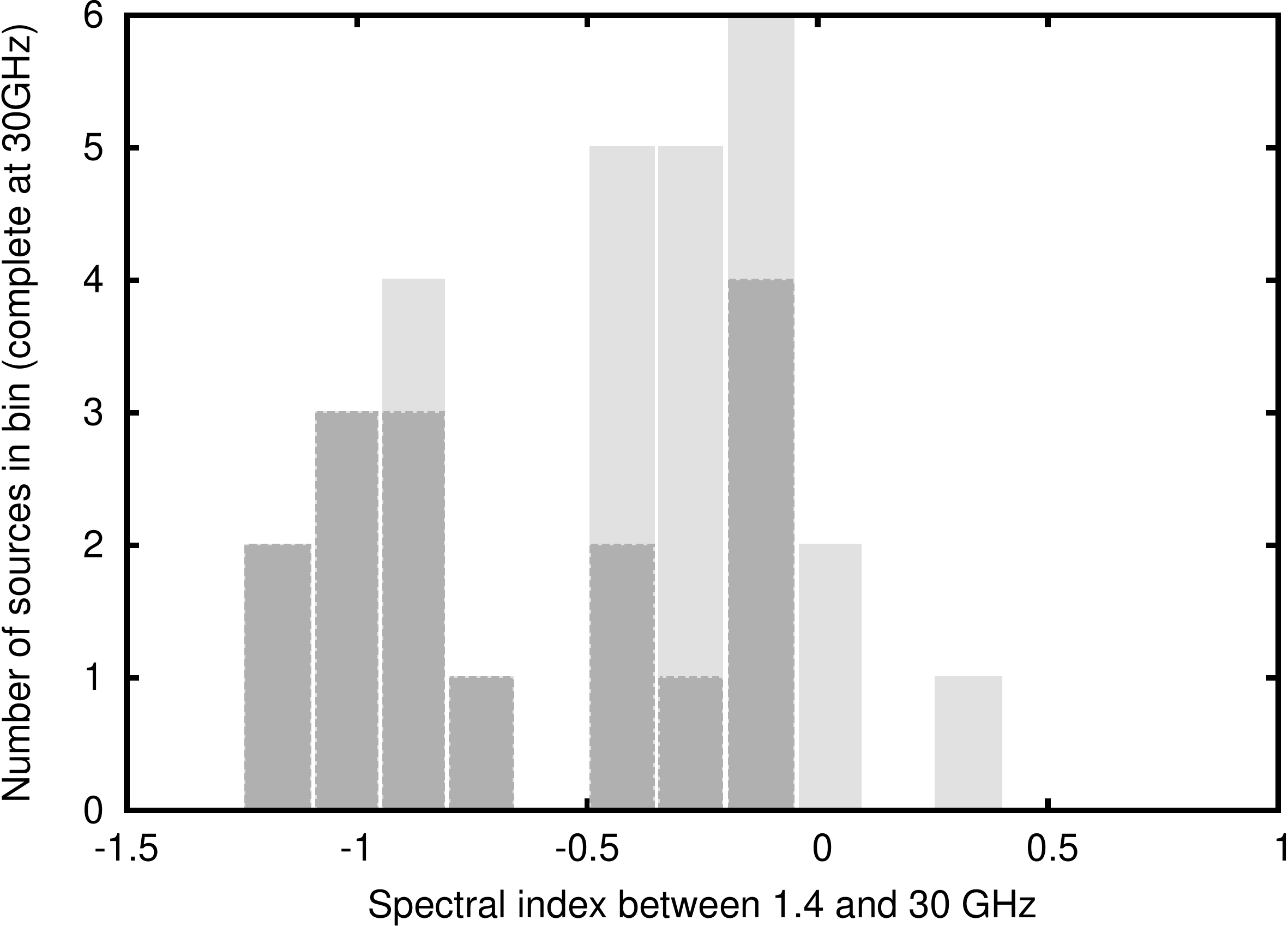}
\caption{The distribution of spectral indices between 1.4 and 30~GHz ($\alpha_{1.4}^{30}$) for 29 sources in this sample with a 30~GHz flux density greater than 10~mJy. The darker boxes represent sources with flux density between 10 and 20~mJy at 30~GHz; the lighter boxes represent the sources with greater than 20~mJy at 30~GHz. The distribution of the stronger sources closely resembles that of the WMAP sample.}
\label{fig:1p4to30_30}
\end{figure}

\begin{figure}
\centering
\includegraphics[scale=0.34]{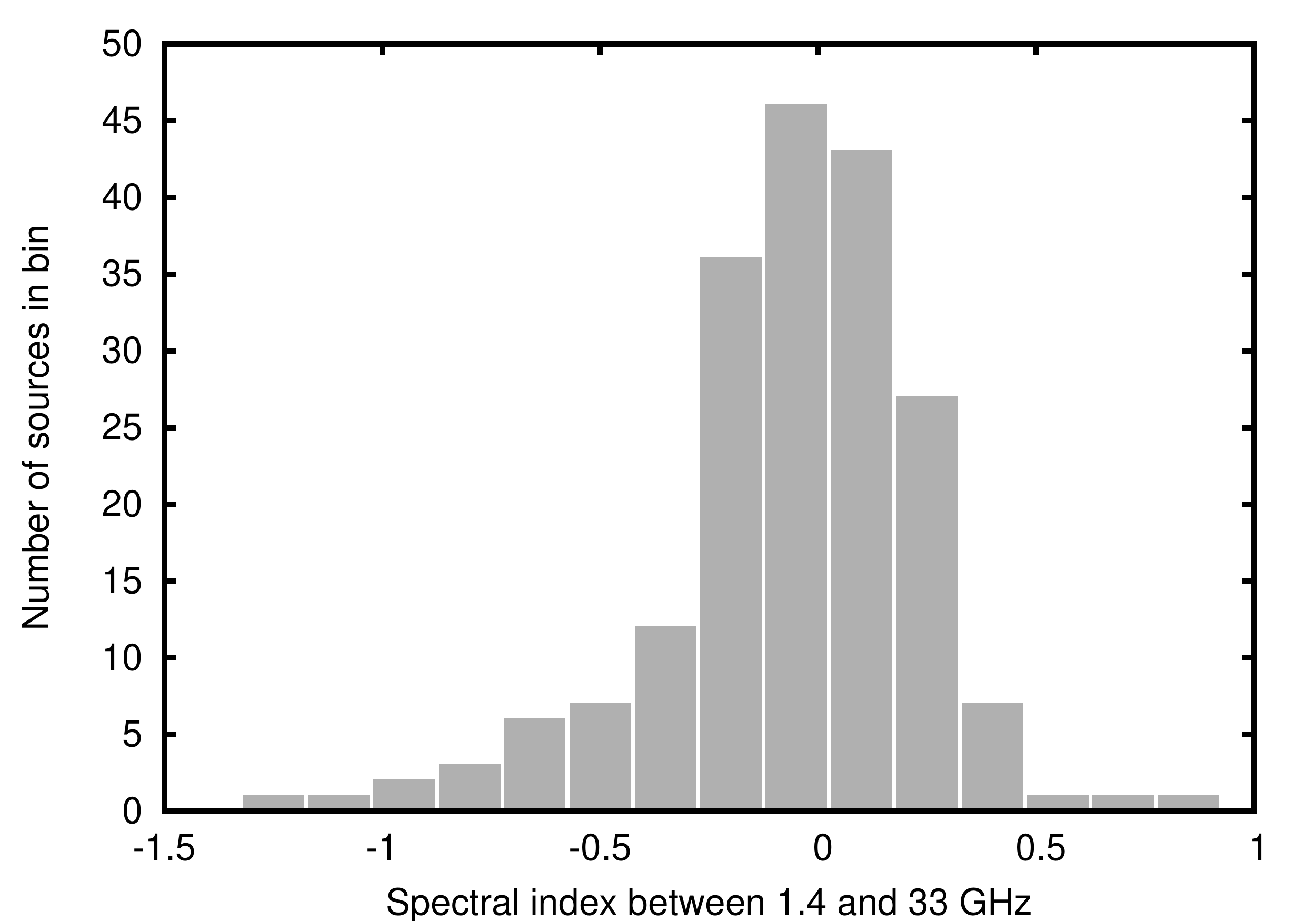}
\caption{The spectral index distribution between 1.4 and 33~GHz of WMAP sources selected to be stronger than 1 Jansky at 33~GHz. The distribution is essentially the same as for the 22~GHz selected sample, but with slightly fewer sources with spectral indices between 0 and -1.}
\label{fig:wmap_indices_30}
\end{figure}

The bimodality of the spectral index distribution remains the same when the sources in this sample are selected at 30~GHz. Figure \ref{fig:1p4to30_30} shows the spectral index distribution of the VSA sources selected to be complete at 30~GHz ($S_{30} > 10$~mJy), and Figure \ref{fig:wmap_indices_30} shows the spectral indices of WMAP sources selected to be greater than 1~Jy at 33~GHz. Splitting the VSA source sample complete at 30~GHz into high ($>$20~mJy) and low ($<$20~mJy) samples shows that the higher flux density sources have a spectral index distribution closer to the WMAP distribution; the lower flux density sources become increasingly more steep spectrum. A Kolmogorov-Smirnov comparison of the VSA ($S_{30} > 10$~mJy) and WMAP distributions shows that the samples are different at the 95 per cent confidence level. The semi-empirical model described by \citet{deZotti2005} predicts a cross-over between the dominance of flat and steep spectrum sources at about 30~mJy (their Figure 14). Our observations display a sharper cross-over than the models appear to suggest, and at the slightly lower flux density of 20~mJy, but are otherwise consistent with the predictions by \citet{deZotti2005}.

\subsection{Morphological properties} \label{sec:morph}
Three of the VSA fields are in the region covered by the FIRST survey \citep{White1997}, and thus 5~arcsec resolution 1.4~GHz maps are available. From a visual inspection of these sources, with contours plotted down to 4 per cent of the peak, we find that 41 of the 68 independent sources (60 per cent) show extended structure; these are marked with an `e' in Table \ref{tab:sourcelist}. Of these sources, the majority have steep spectra ($\alpha_{1.4}^{30}<-0.5$).

Sixty two sources in the sample of WMAP sources described in the previous section also have FIRST maps available. Again we made a visual inspection of maps with the contours drawn to 4 per cent of the peak flux density. We find that only 21 per cent of these sources are extended, a factor three fewer than in the VSA sample. This difference in the proportion of extended sources between samples containing weak and strong sources reinforces our conclusion, based on spectral indices alone, that the mixture of source types is different in the two samples. It is important to consider if this could arise from differences in selection other than the flux density limit. There are several factors which might bias the content of the samples, often in opposite directions:

\begin{itemize}

\item The VSA sources are selected at a lower frequency and therefore there will be a small bias towards more steep spectrum extended sources compared to WMAP.

\item The resolutions of the surveys are very different; the RT has a beam of $\sim$30~arcsec at 15~GHz, whilst WMAP has a beam of $0.88^\circ$ at 22~GHz. The RT observations will discriminate against sources with structures $\gse$~30~arcsec i.e. against very extended steep spectrum sources because emission will be resolved out and the sources will drop below the selection limit.

\item Some VSA sources are weak and thus in several of the FIRST images extended structure would be more difficult to detect than in the much stronger WMAP sources.

\item We do not know the redshift distribution of the VSA sources but being weaker they are likely to be at higher redshift on average than the WMAP sources. This could favour detection of extended structure in the WMAP sources since lower redshift sources will have larger angular sizes and the K-correction will favour higher surface brightness.

\end{itemize}

On balance we think the known selection effects are slightly biased towards finding more extended WMAP sources. We conclude that the population mix really does depend on flux density.

Is the result unexpected? The magnitude of the effect may be larger than expected but not the sign. There is evidence in the data presented in \citet{Orr1982} that at 5~GHz the proportion of flat spectrum (and therefore compact) sources decreases with flux density and they show that this is a robust prediction of relativistic beaming models.  A clearer observational demonstration of the change of population with flux density can be found in \citet{Condon1984}. Condon's Figures 8 and 9 show that the compact flat spectrum source counts at 5~GHz fall off faster with decreasing flux density than the counts of extended sources. This trend probably does not continue down to $\mu$Jy levels since results on the Hubble Deep Field \citep[e.g.][]{Muxlow2005} indicate that compact AGN still form about 25 per cent of the population which at these flux density levels is dominated by star-forming galaxies. For studies of the CMB, the exact population mix is not particularly relevant for total intensity observations, however it may well be important for polarized observations since extended AGNs are likely to have higher linear polarizations than their compact counterparts.

In summary, the above results indicate a marked decrease in the proportion of flat-spectrum sources as a function of decreasing flux density. Such a change in population mix may be important for the accurate prediction of polarized foregrounds for CMB experiments.

\subsection{Comparison of estimates of source surface densities at 30~GHz}

There are 31 sources in the five VSA fields detected above 10~mJy at 30~GHz. These effectively comprise a complete sample as the sources were selected from a deeper survey at the nearby frequency of 15~GHz. The surface density of sources is thus $2 \pm 0.4$ sources per square degree (using the Poissonian error), which can be compared with the extrapolated value of $2.65 \pm 0.1$ in \citet{Mason2009}. \citet{Coble2007} also give an estimate of the surface density of 30~GHz sources in the field, which gives $2.2^{+2.5}_{-1.2}$ sources per square degree; they also point out that the density of sources in clusters of galaxies is significantly higher.

\section{Conclusion}

In order to aid the subtraction of individual sources from the VSA fields observed at $\sim$30~GHz, and to obtain a statistical estimate of the surface density of sources at 30~GHz, we have observed a sample of 121 sources using the OCRA-p receiver on the Toru\'n 32~m telescope; the sample was selected at 15~GHz with the RT. At 30~GHz, we detected 57 sources above a limiting flux density of $\sim 5$~mJy. This is the deepest follow-up of any complete sample of sources detected at 15~GHz by the RT.

At a flux density of 10~mJy, which is our estimated completeness limit, we derive a surface density of sources at 30~GHz of $2.0~\pm~0.4$ per square degree. This is consistent with the value obtained by \citet{Mason2009}, who observed a much larger sample of sources down to mJy levels but selected at a much lower frequency (1.4~GHz). The potential danger of using low frequency selected samples is that there may exist a significant population of sources with steeply rising spectra towards high frequencies that are not present in the low frequency surveys. As the two surface density estimates are consistent, this indicates that such a population is not obviously present at the 10~mJy level.

We have compared our flux density measurements with those from the VSA source subtractor and VLA measurements. These comparisons give confidence in our flux scale but reveal that a significant fraction of sources are variable on a timescale of a few years, some at the level of a factor of 2. This shows the importance of taking contemporaneous measurements of discrete sources in conjunction with measurements of the CMB.

We have also investigated the dependence of the spectral index distribution on flux density by comparing our measured spectral index distribution with that for much stronger sources (above 1~Jy) selected from the WMAP 22 GHz catalogue. We conclude that the proportion of steep spectrum sources increases with decreasing flux density. This is qualitatively consistent with models of source populations, for example \citet{deZotti2005}.

\section*{Acknowledgments}
We gratefully acknowledge the financial support of the Royal Society Paul Instrument Fund which allowed us to construct the 30~GHz OCRA-p receiver. We are also grateful to the Polish Ministry of Science and Higher Education (grant number N N203 390434). M. Peel, M. Davies and T. Franzen acknowledge the support of STFC studentships. This research has made use of the NASA/IPAC Extragalactic Database (NED) which is operated by the Jet Propulsion Laboratory, California Institute of Technology, under contract with the National Aeronautics and Space Administration.

\appendix

\section*{Appendix: Notes on individual sources}
\begin{itemize}
\item J0014+2852a/b -- probably two components of a single source separated by 75~arcsec and our listed 30~GHz flux densities are blends of both components. Not used in the spectral analysis.
\item J0301+2541/42 -- are separated by $\leq$1~arcmin and are probably part of the same source. Their flux densities are blended in our 30~GHz measurements. Not used in the spectral analysis.
\item J0304+2659 -- 15~GHz position is $\sim 15$~arcsec south of NVSS peak.
\item J0936+3129 -- Double source in FIRST map, RT source coincident with western component.
\item J0941+3126 -- the FIRST and NVSS maps show that this source has two extended components. This source is not used in the spectral analysis. 
\item J0941+3226 -- NVSS and FIRST show 15~GHz component is at the southern end of an extended source.
\item J1235+5317 -- 15~GHz position is $\sim 55$~arcsec south-east of NVSS peak.
\item J1235+5311 -- no 1.4~GHz sources are detected in the NVSS or FIRST maps at the RT position but weak extended components straddle this position in the NVSS map. 
\item J1238+5249a/b -- are separated by approximately 2.5~arcmin east-west and thus for half the on-off observations of each source the other lies within the reference beam. The correction, however, is very small.
\item J1535+4142 and J1535+4143a/b -- lie within 3 arcmin of each other. Examination of the FIRST map of the region suggests that they may be physically related. Our 30-GHz flux density measurements are confused. Not used in the spectral analysis.
\item J1541+4114 -- NVSS and FIRST maps show that 15~GHz component is probably the central component of an asymmetric triple source.
\end{itemize}

\bsp

\label{lastpage}

\end{document}